\documentclass{article}

\usepackage{PRIMEarxiv}

\usepackage[utf8]{inputenc} % allow utf-8 input
\usepackage[T1]{fontenc}    % use 8-bit T1 fonts
\usepackage{hyperref}       % hyperlinks
\usepackage{url}            % simple URL typesetting
\usepackage{booktabs}       % professional-quality tables
\usepackage{amsfonts}       % blackboard math symbols
\usepackage{nicefrac}       % compact symbols for 1/2, etc.
\usepackage{microtype}      % microtypography
\usepackage{lipsum}
\usepackage{algorithm}
\usepackage{amsmath}
\usepackage{subfigure}
\usepackage{titlesec}
\usepackage{multirow}
\usepackage{makecell}
\usepackage{algorithm}
\usepackage{algpseudocode}
\usepackage{fancyhdr}       % header
\usepackage{graphicx}       % graphics
\graphicspath{{media/}}     % organize your images and other figures under media/ folder
\usepackage{setspace}
\usepackage{caption}

\usepackage{comment}
\usepackage{footnote}

\usepackage{xcolor}

\title{Machine Learning Methods for Pricing Financial Derivatives}

\author{
  Lei Fan\thanks{This paper is based upon Lei Fan's PhD Thesis at the University of Illinois at Urbana Champaign.} \phantom{..} and Justin Sirignano\thanks{Mathematical Institute, University of Oxford} \\ }

\begin{document}
\maketitle

\begin{abstract}
Stochastic differential equation (SDE) models are the foundation for pricing and hedging financial derivatives. The drift and volatility functions in SDE models are typically chosen to be algebraic functions with a small number ($<5$) parameters which can be calibrated to market data. A more flexible approach is to use neural networks to model the drift and volatility functions, which provides more degrees-of-freedom to match observed market data. Training of models requires optimizing over an SDE, which is computationally challenging. For European options, we develop a fast stochastic gradient descent (SGD) algorithm for training the neural network-SDE model. Our SGD algorithm uses two independent SDE paths to obtain an unbiased estimate of the direction of steepest descent. For American options, we optimize over the corresponding Kolmogorov partial differential equation (PDE). The neural network appears as coefficient functions in the PDE. Models are trained on large datasets (many contracts), requiring either large simulations (many Monte Carlo samples for the stock price paths) or large numbers of PDEs (a PDE must be solved for each contract). Numerical results are presented for real market data including S\&P 500 index options, S\&P 100 index options, and single-stock American options. The neural-network-based SDE models are compared against the Black-Scholes model, the Dupire's local volatility model, and the Heston model. Models are evaluated in terms of how accurate they are at pricing out-of-sample financial derivatives, which is a core task in derivative pricing at financial institutions. Specifically, we calibrate a neural network-SDE model to market data for a financial derivative on an asset with price $S_t$ with a payoff function $g(s)$, and we then evaluate its generalization accuracy for a financial derivative on the same asset $S_t$ but with a different payoff function $f(s)$. In addition to comparing out-of-sample pricing accuracy, we evaluate the hedging performance of the neural network-SDE model.

\end{abstract}


\section{Neural-Network SDE Models} \label{MLoptions}

We will consider the following class of SDE models:
\begin{eqnarray}
 d S_t &=& \mu(S_t, Y_t; \theta) dt + \sigma(S_t, Y_t; \theta) d W_t, \notag \\
 d Y_t &=& \mu_Y(S_t, Y_t; \theta) dt + \sigma_Y(S_t, Y_t; \theta) d Z_t,
 \label{NeuralNetworkSDE}
\end{eqnarray}
where $W_t$ and $Z_t$ are Brownian motions with correlation $\rho$, the initial conditions are $S_0 = s_0$ and $Y_0 = y_0$, and $\mu$, $\sigma$, $\mu_Y, \sigma_Y$ are neural networks (NN) with parameters $\theta$. $S_t$ models the price of the asset (e.g., the stock price). (\ref{NeuralNetworkSDE}) is similar to a stochastic volatility model, except with more degrees-of-freedom due to modeling the drift and volatility functions with machine learning models. The neural network coefficient functions should allow the neural-network SDE (\ref{NeuralNetworkSDE}) to more accurately price complex financial derivatives. 

The parameters $\theta$ must be calibrated to market price data for financial derivatives on the underlying financial asset $S_t$.\footnote{The correlation $\rho$ can also be included in the set of parameters $\theta$.} Once calibrated, the (\ref{NeuralNetworkSDE}) can be used to price and hedge new financial derivatives on the same stock  (i.e., with different maturities, strikes, or payoff functions). 

\subsection{Pricing European Options} \label{EuropeanOption}

Suppose the market prices $P_i^{\textrm{market}}$ for the financial derivatives $i = 1, 2, \ldots, N$ are observed. The maturities and (European) payoff functions for these contracts are $T_i$ and $g_i(s)$, respectively. The buyer of the financial derivative $i$ will receive a (random) payoff of $g_i(S_{T_i})$ at the maturity $T_i$ from the seller. The price of the financial derivative will therefore only depend upon the value of the underlying stock at the final maturity time $T_i$. For a fixed set of parameters $\theta$, the model-generated prices from (\ref{NeuralNetworkSDE}) are:
\begin{eqnarray}
P_i(\theta) = e^{-r T_i} \mathbb{E}[g_i(S_{T_i})].
\label{EuropeanPrice}
\end{eqnarray}
We would like to select the parameter $\theta$ such that the model-generated prices $P_i(\theta)$ match the corresponding market prices $P_i^{\textrm{market}}$ as closely as possible. The objective function for training the SDE model (\ref{NeuralNetworkSDE}) therefore is:
\begin{eqnarray}
J(\theta) = \frac{1}{N} \sum_{i=1}^N \bigg{(} P_i^{\textrm{market}} - P_i(\theta) \bigg{)}^2.
\label{EuropeanOptionObjective}
\end{eqnarray}
Minimizing $J(\theta)$ requires optimizing over the SDE (\ref{NeuralNetworkSDE}), which can be a computationally challenging problem. Furthermore, the stochastic gradient algorithm requires an unbiased estimate of the gradient $\nabla_{\theta} J(\theta)$. We derive a method for sampling an unbiased, stochastic estimate of the gradient $\nabla_{\theta} J(\theta)$ in Section \ref{EuropeanOptimization}.

Once the SDE model has been calibrated to observed market price data by minimizing the objective function (\ref{EuropeanOptionObjective}), it can be used to price new financial derivatives on the stock $S_t$ with different payoff functions and maturities than in the training set. That is, we can generate prices $P_i(\theta)$ for financial derivatives $i = N+1, N+2, \ldots$ with payoff functions $g_{N+1}(s), g_{N+2}(s), \ldots$ and maturities $T_{N+1}, T_{N+2}, \ldots$. Out-of-sample pricing and hedging of financial derivatives which are not liquidly-traded in the market is one of the core applications in quantitative finance. From a machine learning perspective, the goal is to generalize from the observed prices of a series of financial contracts to model the prices of new, different financial contracts for which there are no observed market prices. 

\subsection{Pricing Bermudan and American Options} \label{PricingAmericanOptions}

Financial derivatives with European-style payoff functions only depend upon the value of the underlying financial asset at the final maturity time. However, many financial derivatives depend upon the full path of the underlying stock $S_t$ for $t \in [0, T]$ where $T$ is the final maturity time. An example is a Bermudan option where the option buyer can choose to exercise the option at any time $t \in \{t_0, t_{\Delta}, t_{2 \Delta}, \ldots, t_{M \Delta} \}$ and will receive a payoff $g(S_t)$. An American option allows the option buyer to choose to exercise the option at any time $t \in [0,T]$, which is equivalent to a Bermudan option where $\Delta \rightarrow 0$ and $M = \frac{T}{\Delta}$.

By dynamic programming, the pricing function $P_i(\theta)$ in the objective function will therefore become
\begin{eqnarray}
V^{(m)}(s,y) &=& \mathbb{E} \bigg{[} e^{-r \Delta} \max \big{(} V^{(m+1)}(S_{t_{m+1}}, Y_{t_{m+1}}), g_i(S_{t_m} ) \big{)} \bigg{|} S_{t_m} = s, Y_{t_m} = y  \bigg{]}, \notag \\
P_i(\theta) &=& V^{(0)}(s_0,y_0),
\label{BermudanEqn}
\end{eqnarray}
where $t_m = m \Delta$. Unlike in the case of European options, a stochastic gradient descent algorithm cannot be easily developed to optimize over the objective function (\ref{EuropeanOptionObjective}) due to the nonlinearity of the pricing equation (\ref{BermudanEqn}). Instead, a partial differential equation (PDE) must be solved to price Bermudan or American options. Consequently, in order to calibrate the neural-network SDE model to market prices, a series of PDEs for each Bermudan/American option contract $i = 1, \ldots, N$ must be solved and then optimized over. Typically, there are many contracts on the underlying financial asset and therefore it is a computationally challenging optimization problem. We will use automatic differentiation (AD) and GPUs for fast, parallelized optimization over the PDEs for large numbers of contracts.

\subsection{Literature Review}
Option pricing is a core research area in quantitative finance. Numerous articles have explored theoretical pricing models with different assumptions. In 1973, Black, Scholes, and Merton \cite{black1973pricing, merton1973theory} developed an analytical pricing formula for European options under risk-neutral pricing and the assumption that the underlying asset price follows a Geometric Brownian process. 
 
%The Black-Scholes formula has been historically widely used in options trading \cite{mackenzie2008engine}. 

Many of the assumptions for the Black-Scholes model do not match empirical data. For example, researchers found that the smiles (skewness) and term structures in observed implied volatilities could not be explained with the lognormal assumption of asset price returns. This has motivated the development of more complex SDE models to provide a more accurate fit of market prices. Examples include local volatility models \cite{derman1998stochastic, davis2011dupire}, stochastic volatility models \cite{stein1991stock, heston1993closed, herzel2000option}, and jump-diffusion SDEs \cite{merton1976option} \cite{kou2002jump}.

The first attempt to use neural networks for option pricing and hedging was in the 1990s \cite{malliaris1993beating, malliaris1993neural, hutchinson1994nonparametric, kelly1994valuing, carverhill2003alternative, healy2002data}. These articles directly train a neural network to learn the price of options; that is, neural networks are used as a classical regression model. This is very different than the approach in this paper, where a neural network models drift and volatility functions in an SDE model. The neural networks are then trained by optimizing over the SDE or a corresponding PDE. 

More recently, Funahashi \cite{funahashi2019artificial} presented a new approach to option pricing combining artificial neural networks and an asymptotic expansion. Specifically, an approximation for the option price is obtained by calculating a truncated sum of iterated Ito stochastic integrals. Then an artificial neural network is used to learn the residual between the approximation and the actual value.

In the recent paper \cite{gierjatowicz2020robust}, the authors proposed an approach that combines neural networks with classical SDEs. The neural networks were used to replace coefficient functions within the SDE, to improve the accuracy of the overall pricing model. Their approach is commonly described as a ``neural SDE". The authors present algorithms to calibrate neural SDEs, as well as propose an algorithm to find robust price bounds for an illiquid financial derivative. We also study a class of neural SDE models. However, we develop different optimization methods to train our neural SDE model. We also train and evaluate the neural SDE model on real market data. 

Cohen et al. \cite{cohen2021arbitrage} derived a static arbitrage-free state space for option prices and used neural networks as function approximators for the drift and diffusion of the modelled SDE system. They imposed constraints on the neural nets to preserve the no-arbitrage conditions.
    
\cite{goswami2020data} directly models option prices by carefully selecting features without using historical or implied volatility. There is no SDE in this approach. XGBoost and neural networks are explored for pricing European-style call options. The input features consist of log returns, time to maturity, interest rate, previous option prices, and the covariance matrix of the opening, high, low, and closing price of the underlying asset.

\subsection{Organization of Article}

In Section 3, we derive a new unbiased stochastic gradient descent algorithm that can efficiently optimize over neural-network SDEs to train the neural network parameters. Section 4 implements and evaluates neural-network SDE models on European option datasets, demonstrating that the neural-network-based models have improved pricing accuracy in comparison to a number of traditional mathematical options pricing models. In Section 5, we derive a PDE model for option prices from the neural-network SDE. The PDE model, which is a Kolmogorov backward equation (KBE), can be used to calibrate the neural-network SDE. We optimize over a finite-difference equation approximating the PDE using automatic differentiation (AD). The PDE model allows for calibration using a much wider class of objective functions than the stochastic gradient descent approach. Furthermore, the PDE model allows for calibration on American options, while the stochastic gradient descent approach does not. The method is numerically evaluated on American option datasets. Section 6 evaluates the hedging performance for the neural-network SDE models. 

\section{Optimization} \label{Optimization}

We develop two training methods for the neural-network SDE model. For European options and a mean-squared error objective function, we derive an unbiased stochastic gradient descent algorithm which allows for rapid training of SDE models (Section \ref{EuropeanOptimization}). For general objective functions for European options, we apply a PDE optimization approach (Section \ref{PDEApproach}). For Bermudan and American options, the PDE optimization method is required (Section \ref{OptimizationAmerican}). 

\subsection{Optimization for European Options}  \label{EuropeanOptimization}

The price of $P_i(\theta)$ of the $i$-th financial derivative can be approximated via Monte Carlo simulation of the SDEs $(S_t, Y_t)$ in equation (\ref{NeuralNetworkSDE}):

\begin{eqnarray}
P_i(\theta) \approx P_i^L(\theta) = e^{-r T} \frac{1}{L} \sum_{\ell=1}^L g_i(S_{T}^{\ell}),
\label{EuropeanPrice}
\end{eqnarray}
where $(S_t^{\ell}, Y_t^{\ell})$ are i.i.d. Monte Carlo paths of the SDE $(S_t, Y_t)$ and (without loss of generality) we have set $T_i = T$ for notational convenience. The Monte Carlo approximation $P_i^L(\theta)$ can be used to approximate the objective function $J(\theta)$ as:
\begin{eqnarray}
J(\theta) \approx J^L(\theta) = \frac{1}{N} \sum_{i=1}^N \bigg{(} P_i^{\textrm{market}} - P_i^L(\theta) \bigg{)}^2.
\label{EuropeanOptionObjectiveMC}
\end{eqnarray}
A naive approach would then evaluate the gradient of $J^L(\theta)$ using automatic differentiation and then take a stochastic gradient descent step for the parameters $\theta$. However, this approach is not mathematically correct: it leads to a biased estimate for the gradient. 

In particular, 
\begin{eqnarray}
 \nabla_{\theta} J^L(\theta) &=& -\frac{2}{N} \sum_{i=1}^N \bigg{(} P_i^{\textrm{market}} - P_i^L(\theta) \bigg{)} \nabla_{\theta} P_i^L(\theta) \notag \\
&=& -\frac{2}{N} \sum_{i=1}^N \bigg{(} P_i^{\textrm{market}} - e^{-r T} \frac{1}{L} \sum_{\ell=1}^L g_i(S_T^{\ell}) \bigg{)} \times \nabla_{\theta} \bigg{[} e^{-r T} \frac{1}{L} \sum_{\ell=1}^L g_i(S_T^{\ell}) \bigg{]}.
\label{BiasedGradient}
\end{eqnarray}

$\nabla_{\theta} J^L(\theta)$ is an unbiased estimate for the direction of steepest descent $\nabla_{\theta} J(\theta)$ if $\mathbb{E} [ \nabla_{\theta} J^L(\theta) ] = \nabla_{\theta} J(\theta)$. The direction of steepest descent is
\begin{eqnarray}
 \nabla_{\theta} J(\theta) &=& -\frac{1}{N} \sum_{i=1}^N \bigg{(} P_i^{\textrm{market}} - P_i(\theta) \bigg{)}  \nabla_{\theta} P_i(\theta) \notag \\
 &=& -\frac{1}{N} \sum_{i=1}^N \bigg{(} P_i^{\textrm{market}} - e^{-r T} \mathbb{E}[g_i(S_T)] \bigg{)} \times \nabla_{\theta} \bigg{[} e^{-r T} \mathbb{E}[g_i(S_T)] \bigg{]}.
 \label{DirectionSteepestDescent}
\end{eqnarray}

Comparing (\ref{DirectionSteepestDescent}) and (\ref{DirectionSteepestDescent}), we observe that $\mathbb{E} [ \nabla_{\theta} J^L(\theta) ] \neq \nabla_{\theta} J(\theta)$.\footnote{Generally, $\mathbb{E}[f(S_T)\times g(S_T)] \neq \mathbb{E}[f(S_T)] \times \mathbb{E}[ g(S_T)]$.} Therefore, stochastic gradient descent using the update direction $\nabla_{\theta} J^L(\theta)$ may not converge or may to lead sub-optimal results. 

However, (\ref{BiasedGradient}) can be modified to yield an unbiased, Monte Carlo estimate for the direction of steepest descent. Define:

\begin{eqnarray}
G^L(\theta) = -\frac{2}{N} \sum_{i=1}^N \bigg{(} P_i^{\textrm{market}} - e^{-r T} \frac{1}{L} \sum_{\ell=1}^L g_i(S_T^{\ell}) \bigg{)} \times \nabla_{\theta} \bigg{[} e^{-r T} \frac{1}{L} \sum_{\ell=L+1}^{2L} g_i(S_T^{\ell}) \bigg{]}.
\label{UnBiasedGradient}
\end{eqnarray}
The crucial change that makes (\ref{UnBiasedGradient}) an unbiased estimate for $\nabla_{\theta} J(\theta)$ is that the Monte Carlo samples $(S_t^{L+1}, Y_t^{L+1}), \ldots, (S_t^{2 L}, Y_t^{2 L})$ for the term $\nabla_{\theta} \bigg{[} e^{-r T} \frac{1}{L} \sum_{\ell=L+1}^{2L} g_i(S_T^{\ell}) \bigg{]}$ are independent of the Monte Carlo samples $(S_t^{1}, Y_t^{1}), \ldots, (S_t^{L}, Y_t^{L})$ for the term $\bigg{(} P_i^{\textrm{market}} - e^{-r T} \frac{1}{L} \sum_{\ell=1}^L g_i(S_T^{\ell}) \bigg{)}$. Taking an expectation of (\ref{UnBiasedGradient}), using the independence of these two sets of Monte Carlo samples, and assuming that we can interchange derivatives and expectation yields:
\begin{eqnarray}
 \mathbb{E}[  G^L(\theta) ] = \nabla_{\theta} J(\theta).
\end{eqnarray}
Therefore, for any $L \geq 1$, $G^L(\theta)$
is an unbiased (stochastic) estimate of the direction of steepest descent. This allows for computationally efficient optimization of $J(\theta)$ using modest (or even small) mini-batch sizes.  

The stochastic gradient descent algorithm (with the correct update direction) can then be directly implemented as:
\begin{itemize}
\item Generate $2 L$ Monte Carlo samples of the SDE paths $(S_t, Y_t)$ for the parameters $\theta^{(k)}$.
\item Calculate $G^L(\theta^{(k)})$ using the Monte Carlo samples.
\item Update the parameters using a stochastic gradient descent step:
\begin{eqnarray}
 \theta^{(k+1)} = \theta^{(k)} - \alpha^{(k)} G^L(\theta^{(k)},
\end{eqnarray}
where $\alpha^{(k)}$ is the learning rate.
\end{itemize}

A final important element of the Monte Carlo estimate for the gradient (\ref{UnBiasedGradient}) is that evaluation of the gradients for the contracts $i = 1, 2, \ldots, N$ share the same set of Monte Carlo paths. This saves substantial computational cost in comparison to separately simulating $2 L$ Monte Carlo paths for each contract, which would require a total of $N \times 2 L$ Monte Carlo paths for a single optimization iteration. 

\subsection{PDE Approach to Optimization} \label{PDEApproach}

The stochastic gradient descent method in Section \ref{EuropeanOptimization} can only optimize over a specific objective function: the mean-squared error. In order to optimize over more general objective functions, a PDE approach must be used. 

Consider the objective function:
\begin{eqnarray}
J(\theta) = \frac{1}{N} \sum_{i=1}^N \ell\bigg{(} P_i^{\textrm{market}}, P_i(\theta)  \bigg{)},
\label{GeneralObjectiveFunction}
\end{eqnarray}
where $\ell(z,v)$ is a loss function. The model-generated price $P_i(\theta)$ can be evaluated using a Kolmogorov partial differential equation:
\begin{eqnarray}
 -\frac{\partial v_i}{\partial t} &=& \mu(x,y; \theta) \frac{\partial v_i}{\partial x} + \mu_Y(x,y; \theta) \frac{\partial v_i}{\partial y}  \notag \\
 &+& \frac{1}{2} \sigma(x,y; \theta)^2 \frac{\partial^2 v_i}{\partial x^2}  + \rho \sigma(x, y; \theta) \sigma_Y(x,y; \theta) \frac{\partial^2 v_i}{\partial x \partial y} + \frac{1}{2}  \sigma_Y(x,y; \theta)^2  \frac{\partial^2 v_i}{\partial y^2} - rv, 
 \label{EuropeanKolmogorovPDE}
\end{eqnarray}
where the final condition is $v_i(T,s,y) = g_i(s)$ and $P_i(\theta) = v_i(t = 0, s_0, y_0)$. The PDE (\ref{EuropeanKolmogorovPDE}) must also be completed with appropriate boundary conditions. 

In summary, (\ref{EuropeanKolmogorovPDE}) can be solved to evaluate the model-generated price $P_i(\theta) = v_i(t = 0, s_0, y_0)$ for the parameter $\theta$. Therefore, (\ref{EuropeanKolmogorovPDE}) can be numerically solved using finite-difference methods and then optimized over using automatic differentiation to minimize the objective function (\ref{GeneralObjectiveFunction}).

The advantage of this PDE approach is that it can optimize over a large class of objective functions (\ref{GeneralObjectiveFunction}). For Bermudan and American options, which allow early exercise of the option, the PDE approach is necessary. The stochastic gradient method from Section \ref{EuropeanOptimization} cannot be applied to Bermudan and American options since the price is a nonlinear function of the distribution of the SDE (see Section \ref{PricingAmericanOptions}). 

\subsection{Optimization for Bermudan and American Options} \label{OptimizationAmerican}

The price of a Bermudan option is given by the solution of the nonlinear PDE:

\begin{eqnarray}
 -\frac{\partial v_i^{(m)}}{\partial t} &=& \mu(x,y; \theta) \frac{\partial v_i^{(m)}}{\partial x} + \mu_Y(x,y; \theta) \frac{\partial v_i^{(m)}}{\partial y}  + \frac{1}{2} \sigma(x,y; \theta)^2 \frac{\partial^2 v_i^{(m)}}{\partial x^2}  + \rho \sigma(x, y; \theta) \sigma_Y(x,y; \theta) \frac{\partial^2 v_i^{(m)}}{\partial x \partial y} \notag \\
 &+& \frac{1}{2}  \sigma_Y(x,y; \theta)^2  \frac{\partial^2 v_i^{(m)}}{\partial y^2} - rv^{(m)}, \phantom{.....} t \in [ m \Delta, (m+1) \Delta ],
 \label{BermudanOption}
\end{eqnarray}
where $v_i^{(m)}\big{(} (m+1) \Delta, x,y \big{)} = \max \bigg{(} g_i(x), v_i^{(m+1)} \big{(} (m+1) \Delta,x,y \big{)} \bigg{)}$ and $v_i^{(M)}\big{(} M \Delta, x,y \big{)} = g_i(x) $. The model-generated price $P_i(\theta) = v_i^{(0)}(t = 0, s_0, y_0)$ for the parameter $\theta$. The price of an American option can be approximated by letting $\Delta$ become small. 

We can numerically optimize over the objective function (\ref{GeneralObjectiveFunction}) by using a finite-difference method to discretize (\ref{BermudanOption}) and then using automatic differentiation in combination with gradient descent. The numerical solution and optimization of (\ref{BermudanOption}) is parallelized using GPU computing. A PDE must be solved for each of the $N$ financial derivatives and then the gradient must be calculated for each of these PDEs. The solution of the PDEs is parallelized. At each time step, the solution is represented as a three-dimensional tensor (two spatial dimensions and one dimension corresponding to the different financial derivatives $i =1, \ldots, N$). The finite-difference operators are then applied, the neural networks evaluated at each mesh point, and the three-dimensional tensor solution is updated. Automatic differentiation is used to evaluated the gradient and the parameters $\theta$ are trained with a gradient descent method.

\section{Numerical Results: European Options}

We will train and evaluate the neural network-SDE models on real market data for financial derivatives. The dataset in this section will consist of European call and put options on the S\&P 500 Index. In the next section, we will consider European and American options on the S\&P 100 Index as well as American options on individual stocks. The neural network architecture will be a multi-layer fully-connected network with two hidden layers each consisting of $200$ hidden units. The activation function is the softplus function with $\beta = 1$. Models will be trained on the market prices for options and then evaluated out-of-sample for different time periods and different financial derivatives (e.g., different payoff functions for the same underlying stock). 

Training of the models is computationally costly since there are a large number of financial contracts each day (e.g., $\sim 1,600$ for the S\&P 500) and a large number of Monte Carlo simulations (or PDEs) must be simulated. We will focus therefore on training the model and evaluating it for a large number of financial derivatives but on a relatively small set of days. 

The neural network-SDE model will be benchmarked against widely-used existing SDE models such as the Black-Scholes (BS) model \cite{black1973pricing, merton1973theory}, the local volatility (LV) model \cite{derman1998stochastic, davis2011dupire}, and the Heston model \cite{heston1993closed}. Different variations of the neural network-SDE model will also be compared, including: a 1-D neural network local volatility model, a 1-D neural-network SDE model, a 2-D neural-network SDE model where the drift is set to a constant, and a 2-D neural-network SDE model where the drift is also an output of the neural network. 

A series of models will be compared in our numerical analysis. For convenience, the following abbreviations will be used:

\begin{itemize}
    \item \textbf{BS:} The Black-Scholes model.
    \item \textbf{Local Volatility (LV):} The classic local volatility model where the volatility function is calibrated using Dupire's formula \cite{davis2011dupire, dupire1994pricing}. 
    \item \textbf{Heston:} The Heston model, which is a classic two-dimensional stochastic volatility SDE model. 
    \item \textbf{NNLV:} A neural-network local volatility model 
    \textcolor{black}{    
    \begin{equation}
    dS_t = (r-d)S_tdt + \sigma(S_t,t; \theta)S_t dW_t,
    \label{eq:LocalVSDE}
    \end{equation}
    where $\sigma(S_t,t)$ is a neural network with inputs stock price $S_t$ and time $t$. The neural network is directly calibrated from the data using Dupire's formula; optimization over the SDE is not performed.} \textcolor{black}{Let $f$ be a neural network. $C^{NN}$ is the output and $C$ is the true call option prices. We train the neural network on the available call options using the following algorithm.}
    \begin{algorithm}[H] 
    \caption{Optimization algorithm}\label{euclid}
    \begin{algorithmic}[1]
    \Procedure{}{}
    \For{\texttt{$epoch \gets 1$ to $epochs$ }}
    \State $C^{NN} \gets f(S, t)$
    \State $loss \gets MSE(C, C^{NN})$
    % \State $MSELoss \gets \sum((C/S_0 - C_1)^2)$
    \State loss.backward()
    \State optimizer.step()
    \EndFor
    % \State \textbf{goto} \emph{top}.
    \EndProcedure
    \end{algorithmic}
    \end{algorithm}
    \textcolor{black}{After training, we use this neural network $f$ directly for simulation and prediction. Here $g$ denotes Dupire's formula.}
    \begin{algorithm}[H] 
    \caption{Simulation algorithm}\label{euclid}
    \begin{algorithmic}[1]
    \Procedure{}{}
    \For{\texttt{$epoch \gets 1$ to $epochs$ }}
    \State $S_0 \gets ones(N,L)$
          \For{\texttt{$t \gets 0$ to $M-1$ }}
            \State $C_t^{NN} \gets f(S_t, t)$
            \State $\sigma_t \gets g(C_t^{NN}, t)$
            \State $S_{t+1} \gets S_t + (r-d)S_t\Delta + \sigma_t S_t \sqrt{\Delta}Z_t$
          \EndFor
    
    % \State $P \gets \frac{1}{L}\sum_{i=1}^{L} [e^{-rT}max(K/S_0 - S^i , 0)]$
    \State $C \gets \frac{1}{L}\sum_{i=1}^{L} [e^{-rT}max(S^i - K/S_0 , 0)]$
    \State $P \gets \frac{1}{L}\sum_{i=1}^{L} [e^{-rT}max(K/S_0 - S^i , 0)]$
    % \State loss.backward()
    % \State optimizer.step()
    \EndFor
    % \State \textbf{goto} \emph{top}.
    \EndProcedure
    \end{algorithmic}
    \end{algorithm}
    
    \item \textbf{SDENN:} A neural-network SDE model where we optimize over the entire SDE model using a stochastic gradient descent method: 
    \begin{eqnarray}
    dS_t &=& (r-d)S_tdt + \sigma(S_t,t; \theta) S_t dW_t, \notag \\
        \sigma^2(S_t,t; \theta) &=& 2\frac{\frac{\partial C(S_t,t)}{\partial T} + (r-d)K\frac{\partial C(S_t,t; \theta)}{\partial K} + dC(S_t,t; \theta)}{K^2\frac{\partial^2 C(S_t,t; \theta)}{\partial K^2}},
    \label{eq:LocalVSDE}
    \end{eqnarray}
    where $C(K,T; \theta)$ is a neural network with inputs stock price $S_t$ and time $t$. $\sigma(C)$ is the Dupire's formula with the call option price as the input. The functional form of Dupire's formula is applied to the output of the fully-connected neural network and can be considered part of the neural network architecture. 
    
    In contrast to the NNLV model, we optimize over the entire SDE to train the parameters. In the NNLV model, the parameters for the volatility model are calibrated off-line using Dupire's formula and then substituted into the SDE for simulation. Therefore, the SDENN model optimizes over the actual simulation which generates the price while NNLV performs an off-line optimization which may lead to sub-optimal model simulation performance. 
    \item \textbf{SDENN-Drift:}  A neural-network SDE model where we optimize over the entire SDE model using stochastic gradient descent. Both the drift \emph{and} volatility are modeled with neural networks. The SDE model is: 
    \begin{eqnarray}
    dS_t &=& \mu(S_t, t; \theta)S_tdt + \sigma(S_t,t; \theta) S_t dW_t, \notag \\
        \sigma^2(S_t,t; \theta) &=& 2\frac{\frac{\partial C(S_t,t; \theta)}{\partial T} + (r-d)K\frac{\partial C(S_t,t; \theta)}{\partial K} + dC(S_t,t; \theta)}{K^2\frac{\partial^2 C(S_t,t; \theta)}{\partial K^2}},
    \label{eq:LocalVSDE}
    \end{eqnarray}
    where $\mu(S_t,t; \theta)$ and $C(S_t,t; \theta)$ are two neural networks with inputs stock price $S_t$ and time $t$.
    
    \item \textbf{2D-NN:} A \emph{two-dimensional} neural-network SDE model where we optimize over the entire SDE model using stochastic gradient descent. The first SDE models the price while the second SDE models the stochastic volatility. The SDE model is:
    \begin{eqnarray}
        dS_t &=& f_1(S_t, Y_t, t; \theta)dt + f_2(S_t, Y_t, t; \theta) dW_t^S, \\
        dY_t &=& f_3(S_t, Y_t, t; \theta) dt + f_4(S_t, Y_t, t; \theta) dW_t^Y,
    \end{eqnarray}
    where $W_t^S$ and $W_t^Y$ are two standard Brownian motions with correlation $\rho$, $f(s, y, t; \theta)$ is a neural network, and $f_i(s,y,t; \theta)$ is the $i$-th output of the neural network. The correlation $\rho$ and the initial value $Y_0$ are trained together with the neural network parameters $\theta$.
    
\end{itemize}

We present results from several experiments: intraday out-of-sample prediction, the next-day prediction, prediction for out-of-sample payoff functions, and prediction for out-of-sample strike prices. In the first two experiments, the neural-network-based models have lower MSE and MAE than the traditional SDE models. The 2D-NN model achieves the best performance. The latter two experiments demonstrate the ability of the neural networks to accurately price new, \emph{previously unseen} payoff functions. That is, the neural network-SDE model is able to \emph{generalize} to new types of financial derivatives which were not in the training dataset.

\subsection{Intraday Out-of-Sample Prediction}\label{IntradaySDE}

In our first experiment, we train and test on European options on the S\&P 500 Index for the same day. We organize the call and put contracts at the same strike price and maturity into pairs. Then, 80\% of the option pairs are randomly selected for the train set and 20\% of the option pairs are selected for the test set. 

\begin{figure}[h]
    % \raggedleft
    \hspace{0.5cm}\includegraphics[width=0.94\textwidth]{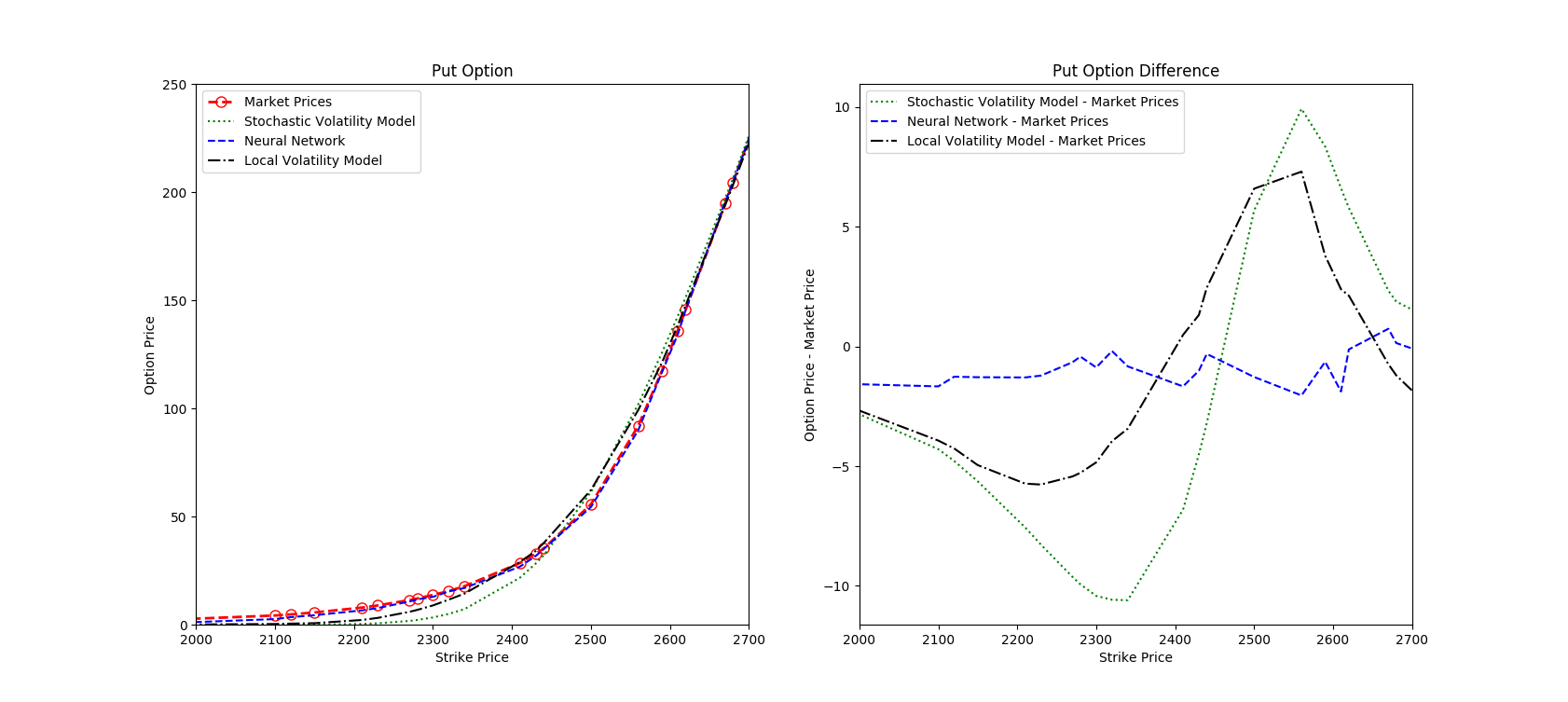}
    \caption{Comparison of the neural-network SDE model, stochastic volatility model, and local volatility model for put options on 2017-09-01. The time to maturity is 77 days. The left figure displays the intraday out-of-sample prediction for the put option prices. The right figure shows the difference between the predicted prices and the actual market prices.}
    \label{fig:Intro}
\end{figure}

Table \ref{Tab:intradaysummary1} compares the out-of-sample performance of the neural network-SDE model against standard SDE models such as Black-Scholes, Local Volatility, and Heston. Table \ref{Tab:intradaysummary2} compares various versions of neural-network SDE models. Models are evaluated according to the mean-squared error and the mean-absolute error averaged across all contracts in the test dataset. Figure \ref{fig:Intro} plots the model-generated prices versus the strike price. We can observe that neural-network-based models perform better than traditional option pricing models, and that the two-dimensional neural network-SDE has the highest out-of-sample accuracy. 

\begin{table}[H]
    \centering
    \begin{tabular}{ |p{1.6cm}<{\centering}|p{1.3cm}<{\centering}|p{1.3cm}<{\centering}|p{1.3cm}<{\centering}|p{1.3cm}<{\centering}|p{1.3cm}<{\centering}|p{1.3cm}<{\centering}|p{1.3cm}<{\centering}|p{1.3cm}<{\centering}| }
     \hline
     \multicolumn{9}{|c|}{Summary of intraday out-of-sample prediction for different models.} \\
     \hline
     Days & \multicolumn{2}{c|}{BS} & \multicolumn{2}{c|}{Local Volatility} & \multicolumn{2}{c|}{Heston} &  \multicolumn{2}{c|}{2D-NN} \\
     \hline
      & call & put & call & put & call & put &  call & put\\
     \hline
     \multirow{2}{*}{2017-09-01}  & 118.41 & 53.87 & 121.92 & 15.24 & 73.47  & 49.18 & 9.51 & 2.90\\
                                  & 102.79\% &  64.76\% & 97.61\% &  60.69\% & 55.45\% & 66.29\% & 8.83\% &  41.10\%\\
     \hline
     \multirow{2}{*}{2017-10-23}  & 128.34 & 92.56 & 148.08 & 30.53 & 182.51  & 79.02 & 7.73 & 3.07\\
                                 & 56.54\% &  69.38\%  & 40.69\% &  64.18\% & 30.30\% & 72.27\% & 5.91\% &  42.84\%\\
     \hline
     \multirow{2}{*}{2017-11-10}  & 123.10 & 90.47 & 111.42 & 37.28 & 70.16  & 74.69 & 6.59 & 2.20\\
                                  & 79.39\% &  67.71\% & 68.33\% &  63.74\% & 53.03\% & 67.88\% & 7.00\% &  40.97\%\\
     \hline
     \multirow{2}{*}{2017-12-07}  & 79.90 & 47.69 & 69.30 & 24.33 & 196.20 & 48.94 & 5.89 & 1.80\\
                                  & 58.31\% &  68.31\%& 46.93\% & 67.73 \% & 40.77\% & 71.21\% & 9.42\% & 41.95\% \\
     \hline
     \multirow{2}{*}{2018-01-26}  & 213.96 & 133.70 & 200.93 & 46.18 & 132.97 & 97.78 & 10.70 & 10.10\\
                                  & 30.12\% &  65.31\% & 20.69\% &  62.43\% & 15.71\% & 65.67\% & 8.20\% &  36.34\% \\
     \hline
    \end{tabular}\\
    \caption{Summary of intraday out-of-sample prediction for different models. The first row is the mean squared error and the second row is the relative mean absolute error (in percent). The formula for the relative mean absolute error is $\frac{1}{N} \sum_{i=1}^N \frac{ |P_i(\theta) - P_i^{\textrm{market}} | }{P_i^{\textrm{market}} } \times 100\%$.}
    \label{Tab:intradaysummary1}
\end{table}

\vspace{-1.5em}

\begin{table}[H]
    \centering
    \begin{tabular}{ |p{1.6cm}<{\centering}|p{1.3cm}<{\centering}|p{1.3cm}<{\centering}|p{1.3cm}<{\centering}|p{1.3cm}<{\centering}|p{1.3cm}<{\centering}|p{1.3cm}<{\centering}|p{1.3cm}<{\centering}|p{1.3cm}<{\centering}| }
     \hline
     \multicolumn{9}{|c|}{Summary of intraday out-of-sample prediction for deep learning models.} \\
     \hline
     Days & \multicolumn{2}{c|}{NNLV} & \multicolumn{2}{c|}{SDENN} & \multicolumn{2}{c|}{SDENN-Drift} &  \multicolumn{2}{c|}{2D-NN} \\
     \hline
      & call & put & call & put & call & put &  call & put\\
     \hline
     \multirow{2}{*}{2017-09-01}  & 93.61 & 14.98 & 96.63 & 15.58 & 26.91 & 6.23 & 9.51 & 2.90\\
                                  & 26.38\% &  49.14\% & 19.13\% &  48.50\% & 32.27\% &  44.95\% & 8.83\% &  41.10\%\\
     \hline
     \multirow{2}{*}{2017-10-23}  & 114.79 & 37.27 & 115.01 & 30.31 & 11.87 & 13.42 & 7.73 & 3.07\\
                                 & 9.39\% &  60.09\%  & 11.97\% &  59.34\% & 11.03\% &  54.49\% & 5.91\% &  42.84\%\\
     \hline
     \multirow{2}{*}{2017-11-10}  & 91.33 & 46.27 & 94.45 & 39.16 & 17.03 & 13.07 & 6.59 & 2.20\\
                                  & 14.16\% &  60.16\% & 14.71\% &  57.75\% & 16.31\% &  49.63\% & 7.00\% &  40.97\%\\
     \hline
     \multirow{2}{*}{2017-12-07}  & 60.57 & 49.48 & 64.04 & 33.02 & 33.88 & 16.73 & 5.89 & 1.80\\
                                  & 12.37\% &  70.75\%& 12.98\% & 68.17\% & 13.73\% &  65.64\% & 9.42\% & 41.95 \% \\
     \hline
     \multirow{2}{*}{2018-01-26}  & 189.98 & 52.85 & 21.88 & 22.39 & 21.88 & 22.39 & 10.70 & 10.10\\
                                  & 14.16\% &  60.16\% & 18.18\% &  59.75\% & 13.26\% &  54.91\% & 8.20\% &  36.34\% \\
     \hline
    \end{tabular}\\
    \caption{Summary of intraday out-of-sample prediction for deep learning models. The first row is the mean squared error and the second row is the mean absolute error rate.}
    \label{Tab:intradaysummary2}
\end{table}

\vspace{-2.5em}

\subsection{The Next-day Prediction} \label{NextDaySDE}

In this section, we train models on day $t$ and then test their out-of-sample performance for pricing financial contracts on day $t+1$. Table \ref{Tab:NextDaysummary1} compares the out-of-sample performance of the neural network-SDE model against standard SDE models such as Black-Scholes, Local Volatility, and Heston. Table \ref{Tab:NextDaysummary2} compares different versions of the neural network-SDE model. We can observe that neural-network-based models have lower MSE and MAE than traditional option pricing models, and that the 2-dimensional neural network-SDE (2D-NN) demonstrated the best performance.

\begin{table}[H]
    \centering
    \begin{tabular}{ |p{1.6cm}<{\centering}|p{1.3cm}<{\centering}|p{1.3cm}<{\centering}|p{1.3cm}<{\centering}|p{1.3cm}<{\centering}|p{1.3cm}<{\centering}|p{1.3cm}<{\centering}|p{1.3cm}<{\centering}|p{1.3cm}<{\centering}| }
     \hline
     \multicolumn{9}{|c|}{Summary of next-day out-of-sample prediction for different models.} \\
     \hline
     Days & \multicolumn{2}{c|}{BS} & \multicolumn{2}{c|}{Local Volatility} & \multicolumn{2}{c|}{Heston} &  \multicolumn{2}{c|}{2D-NN}\\
     \hline
      & call & put & call & put & call & put &  call & put\\
     \hline
     \multirow{2}{*}{2017-09-05} & 488.56 & 172.73 & 330.08 &  73.78 & 291.25 & 142.13 &  121.93 & 52.46 \\
                                 & 110.87\% &  69.29\% & 105.79\% &  64.46\% & 64.84\% & 70.24\% & 11.20\% &  51.70\% \\
     \hline
     \multirow{2}{*}{2017-10-24} & 101.33 & 91.90 & 157.92 & 32.37 & 136.79 & 72.27 & 8.30 & 2.78\\
                                 & 52.33\% &  72.01\% & 58.40\% &  67.29\% & 29.86\% & 73.35\% & 6.65\% &  48.06\%\\
     \hline
     \multirow{2}{*}{2017-11-13} & 88.15 & 84.21 & 54.87 & 39.59 &  68.80 & 74.26 & 40.11 & 9.89\\
                                 & 65.15\% &  68.90\% & 55.05\% &  66.42\% & 42.68\% & 69.48\% & 11.69\% &  43.22\%\\
     \hline
     \multirow{2}{*}{2017-12-08} & 125.17 & 112.01 & 53.90 & 39.41 &  67.09 & 52.40 & 11.78 & 9.21\\
                                 & 49.46\% &  69.79\% & 53.39\% &  66.56\% & 35.33\% & 68.96\% & 8.07\% &  39.39\%\\
     \hline
     \multirow{2}{*}{2018-01-29} & 527.35 & 209.04 & 349.77 & 117.23 & 347.56 & 173.25 & 55.72 & 24.51\\
                                 & 29.53\% &  70.34\% & 18.68\% & 68.62 \% & 17.33\% & 71.50\% & 9.74\% &  45.92\%\\
     \hline
    \end{tabular}\\
    \caption{Summary of next-day out-of-sample predictions for different models. The first row is the mean squared error and the second row is the mean absolute error rate.}
     \label{Tab:NextDaysummary1}
\end{table}

\vspace{-2em}

\begin{table}[H]
    \centering
    \begin{tabular}{ |p{1.6cm}<{\centering}|p{1.3cm}<{\centering}|p{1.3cm}<{\centering}|p{1.3cm}<{\centering}|p{1.3cm}<{\centering}|p{1.3cm}<{\centering}|p{1.3cm}<{\centering}|p{1.3cm}<{\centering}|p{1.3cm}<{\centering}| }
     \hline
     \multicolumn{9}{|c|}{Summary of next-day out-of-sample prediction for deep learning models.} \\
     \hline
     Days & \multicolumn{2}{c|}{NNLV} & \multicolumn{2}{c|}{SDENN} & \multicolumn{2}{c|}{SDENN-Drift} &  \multicolumn{2}{c|}{2D-NN}\\
     \hline
      & call & put & call & put & call & put &  call & put\\
     \hline
     \multirow{2}{*}{2017-09-05} & 305.63 & 106.37 & 321.66  & 92.31 & 169.12 & 34.13 &  121.93 & 52.46 \\
                                 & 34.69\% & 59.33\% & 40.27\% & 57.00\% & 37.25\% &  48.02\% & 11.20\% &  51.70\% \\
     \hline
     \multirow{2}{*}{2017-10-24} & 121.08  & 37.88 & 128.29 & 29.79 & 13.20 & 11.74 & 8.30 & 2.78\\
                                 & 17.96\% & 65.07\% & 26.78\% & 62.26\% & 20.18\% &  58.16\% & 6.65\% &  48.06\%\\
     \hline
     \multirow{2}{*}{2017-11-13} & 42.38 & 26.01 & 40.80 & 20.69 &  39.62 & 21.31 & 40.11 & 9.89\\
                                 & 16.75\% & 59.24\% & 16.94\% & 56.35\% & 20.89\% &  49.27\% & 11.69\% &  43.22\%\\
     \hline
     \multirow{2}{*}{2017-12-08} & 48.33 & 38.69 & 42.30 & 23.21 &  71.02 & 20.21 & 11.78 & 9.21\\
                                 & 11.39\% & 67.34\% & 11.19\% & 64.12\% & 12.20\% &  61.81\% & 8.07\% &  39.39\%\\
     \hline
     \multirow{2}{*}{2018-01-29} & 332.92  & 142.82 & 338.95 & 133.33 & 75.55 & 59.54 & 55.72 & 24.51\\
                                 & 9.78\% & 66.61\% & 11.99\% & 66.04\% & 9.09\% &  61.94\% & 9.74\% &  45.92\%\\
     \hline
    \end{tabular}\\
    \caption{Summary of next-day out-of-sample predictions for deep learning models. The first row is the mean squared error and the second row is the mean absolute error rate.}
    \label{Tab:NextDaysummary2}
\end{table}

\vspace{-2.5em}
\subsection{Out-of-sample Payoff Functions} \label{OutOfSampleC2}
\begin{figure}[H]
    % \raggedleft
    \hspace{-1.5cm}\includegraphics[width=1.15\textwidth]{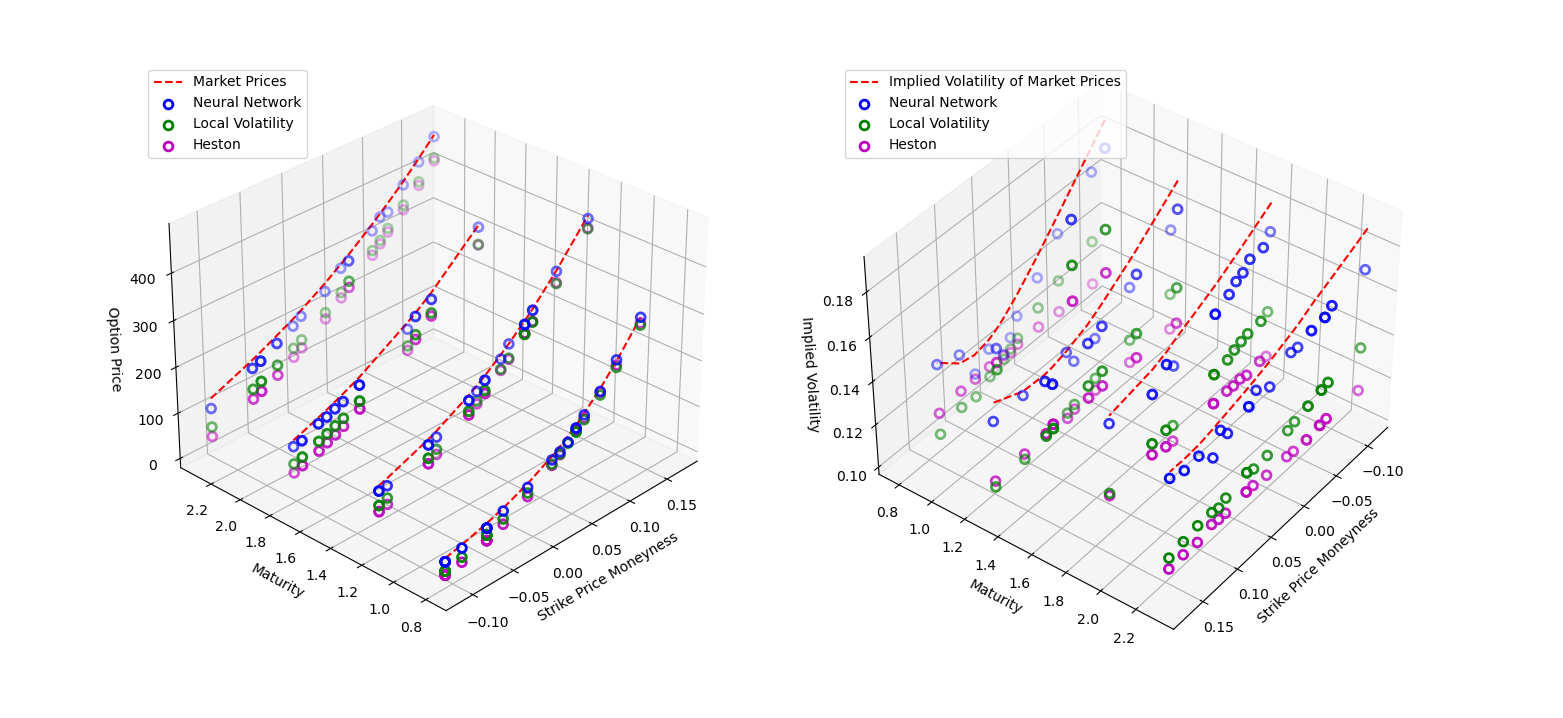}
    \caption{The model-generated option price curve for 2017-09-01 where we trained on the call options and evaluated out-of-sample on the put options. The left figure is the put options curve in which the z-axis is the option price. The right figure is the corresponding implied volatility curve. The results are out-of-sample.}
     \label{figEuropeanPutOption1}
\end{figure}

\begin{figure}[H]
    % \raggedleft
    \hspace{-1.5cm}\includegraphics[width=1.15\textwidth]{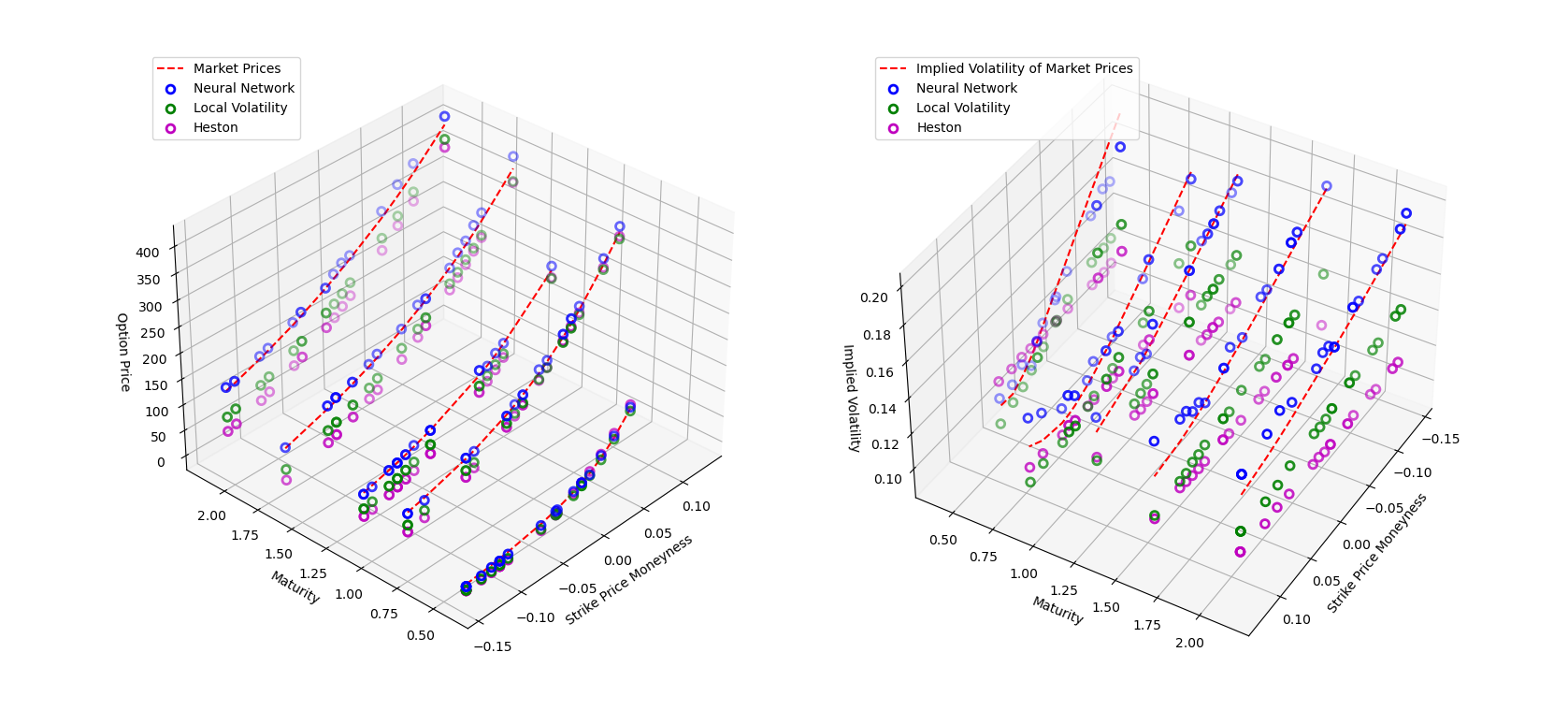}
    \caption{The model-generated option price curve for 2017-10-23 where we trained on the call options and evaluated out-of-sample on the put options. The left figure is the put options curve in which the z-axis is the option price. The right figure is the corresponding implied volatility curve. The results are out-of-sample.}
     \label{figEuropeanPutOption2}
\end{figure}

\begin{figure}[H]
    % \raggedleft
    \hspace{-1.5cm}\includegraphics[width=1.15\textwidth]{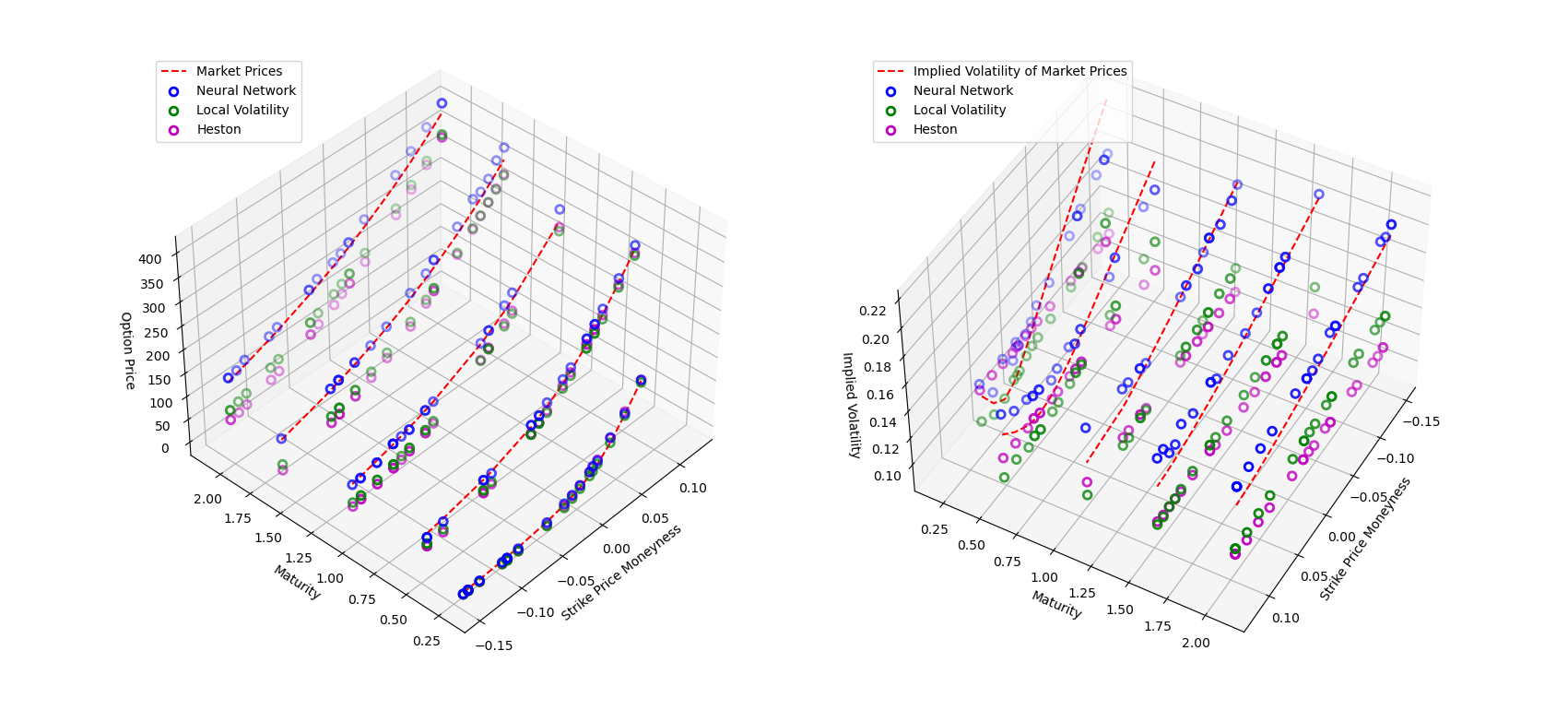}
    \caption{The model-generated option price curve for 2017-11-10 where we trained on the call options and evaluated out-of-sample on the put options. The left figure is the put options curve in which the z-axis is the option price. The right figure is the corresponding implied volatility curve. The results are out-of-sample.}
     \label{figEuropeanPutOption3}
\end{figure}

In this section, we train on call options and then test the model accuracy on put options. Table \ref{Tab:summaryEuropeanTestPayoffFunction} compares the performance of the neural network-SDE model with the benchmark SDE models. Figures \ref{figEuropeanPutOption1}, \ref{figEuropeanPutOption2}, and \ref{figEuropeanPutOption3} compare the accuracy of the neural network-SDE with the local volatility and Heston models across different maturities and strike prices. The neural network-SDE substantially outperforms the Heston and local volatility models. 

\vspace{-0.5em}
\begin{table}[H]
    \centering
    \begin{tabular}{ |p{1.6cm}<{\centering}|p{1.3cm}<{\centering}|p{1.3cm}<{\centering}|p{1.3cm}<{\centering}|p{1.3cm}<{\centering}|p{1.3cm}<{\centering}|p{1.3cm}<{\centering}|p{1.3cm}<{\centering}|p{1.3cm}<{\centering}| }
     \hline
     \multicolumn{9}{|c|}{Train on Call Options and Test on Put Options.} \\
     \hline
     Days & \multicolumn{2}{c|}{BS}  & \multicolumn{2}{c|}{Local Volatility} & \multicolumn{2}{c|}{Heston}  & \multicolumn{2}{c|}{2D-NN} \\
     \hline
      & Call & Put  & Call & Put  & Call & Put  & Call & Put\\
     \hline
     \multirow{3}{*}{2017-09-01} & 11.517 & 11.374 & 10.608 & 9.800  &  11.476 & 11.372 & 3.423 & 4.062\\
                               & 417.277 & 489.360 & 407.999 & 334.586   & 415.468 & 488.180 & 44.431 & 54.009\\
                                & 56.350\% & 70.914\%  & 26.218\% & 68.255\% & 54.958\% & 70.908\% & 10.544\% & 55.979\%\\

    \hline
     \multirow{3}{*}{2017-10-23}& 12.457 & 11.461 & 11.393 & 8.977  & 12.400 & 11.454 & 3.144 & 2.724\\
                                & 356.374 & 421.542 & 334.358 & 219.252 & 356.032 & 420.266 & 24.972 & 18.435\\
                                & 51.314\% & 71.344\%  & 27.876\% & 68.126\%  & 49.688\% & 71.355\% & 8.833\% & 49.297\%\\

    \hline
     \multirow{3}{*}{2017-11-10} & 11.351 & 11.348 & 10.132 & 10.230  & 11.291 & 11.335 & 3.977 & 3.149\\
                                & 328.135 & 451.155 & 306.366 & 318.385  & 325.801 & 448.889 & 35.787 & 29.735\\
                                & 59.638\% & 69.972\% & 52.144\% & 67.729\% & 57.813\% & 69.982\% & 26.553\% & 51.901\%\\

    \hline
    \end{tabular}\\
    \caption{Models are trained on call options and then evaluated out-of-sample on put options. There are three columns for each cell. The first column is for "MAE" (mean absolute error). The second column is for "MSE" (mean squared error). The third column is for "relative MAE" (relative mean absolute error).}
    \label{Tab:summaryEuropeanTestPayoffFunction}
\end{table}
From the table, we observe that the neural-network-based model has the lowest errors, while the other three traditional pricing models have similar performance.

\subsection{Out-of-sample Strike Prices} \label{OutOfSampleSP2}
\begin{figure}[H]
    % \raggedleft
    \hspace{-1.5cm}\includegraphics[width=1.20\textwidth]{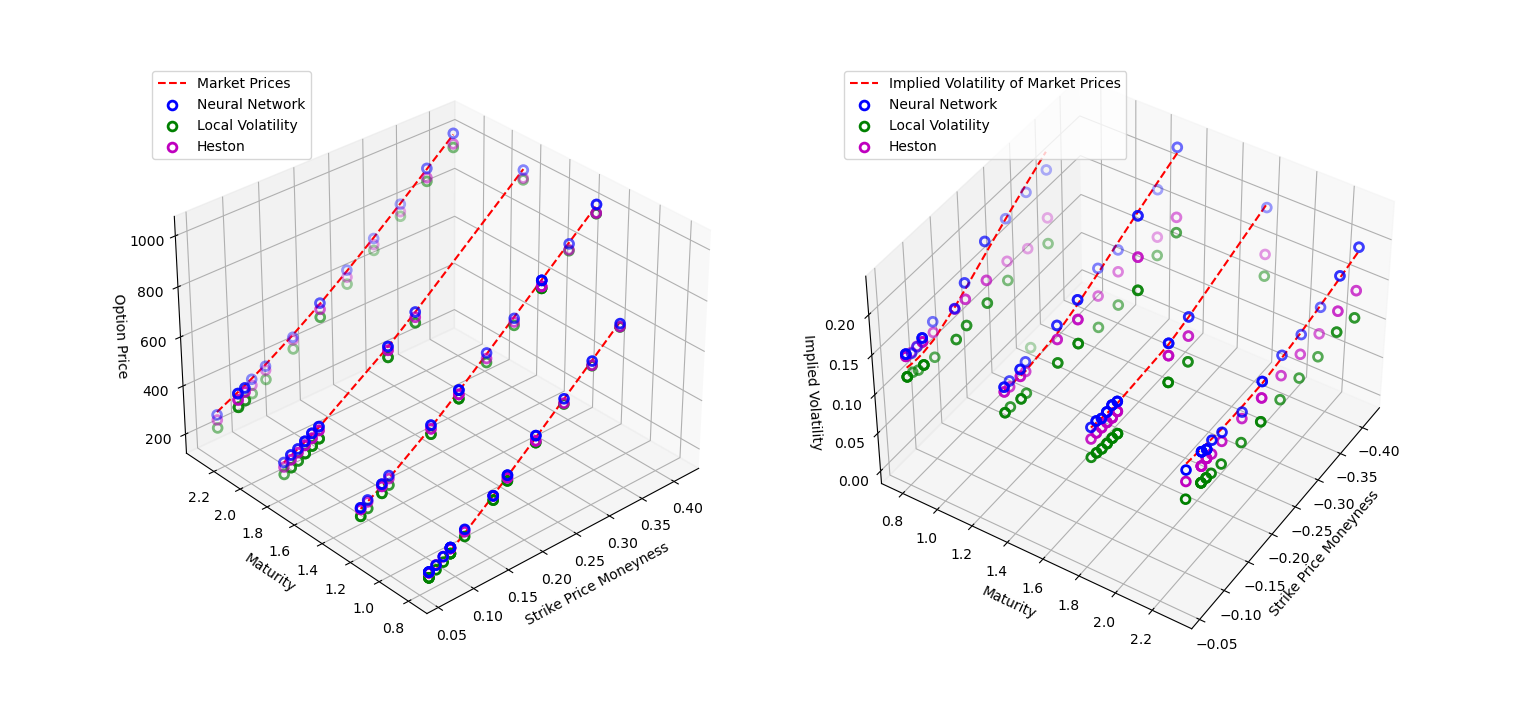}
    \caption{The model-generated option price curve for 2017-09-01 where models are trained on the small strike price contracts and evaluated out-of-sample on the contracts with large strike prices. The left figure is the call option curve and the right figure is the put option curve. The z-axis is the option price.}
     \label{fig:summaryNewStrikePrices1}
\end{figure}

\begin{figure}[H]
    % \raggedleft
    \hspace{-1.5cm}\includegraphics[width=1.20\textwidth]{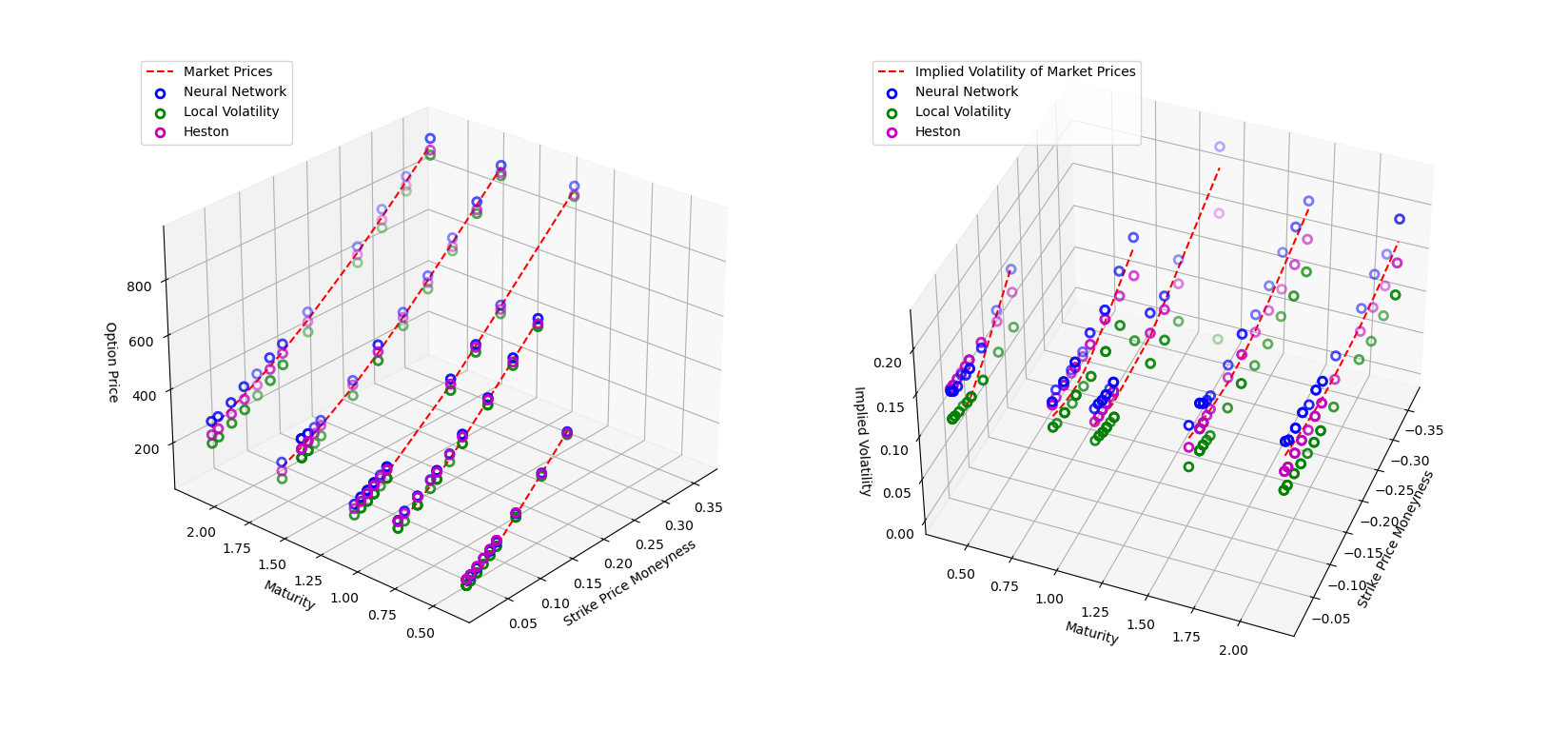}
    \caption{The model-generated option price curve for 2017-10-23 where models are trained on the small strike price contracts and evaluated out-of-sample on the contracts with large strike prices. The left figure is the call option curve and the right figure is the put option curve. The z-axis is the option price.}
    \label{fig:summaryNewStrikePrices2}
\end{figure}

\begin{figure}[H]
    % \raggedleft
    \hspace{-1.5cm}\includegraphics[width=1.20\textwidth]{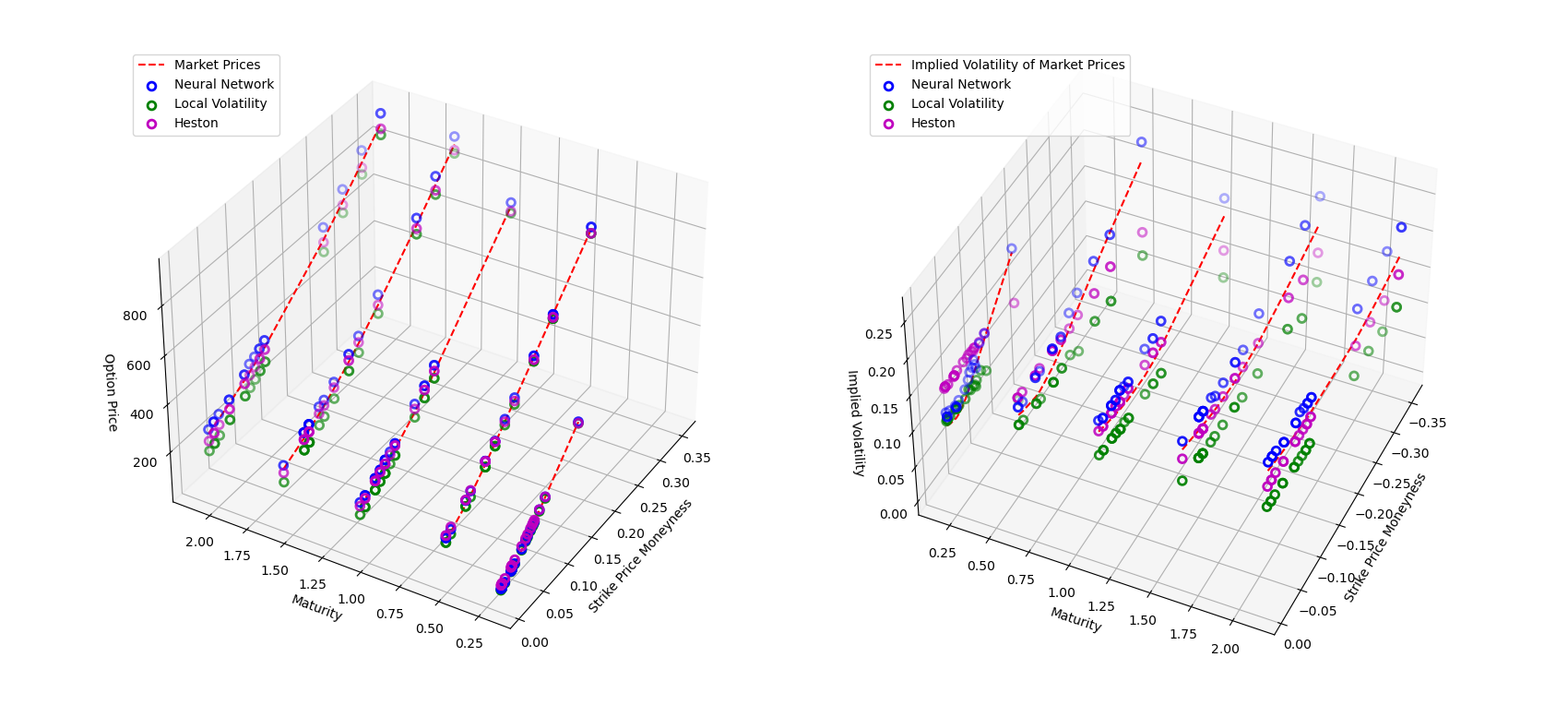}
    \caption{The model-generated option price curve for 2017-11-10 where models are trained on the small strike price contracts and evaluated out-of-sample on the contracts with large strike prices. The left figure is the call option curve and the right figure is the put option curve. The z-axis is the option price.}
    \label{fig:summaryNewStrikePrices3}
\end{figure}

We also test the model performance for out-of-sample strike prices. The analysis in this section is based upon the same train/test sets as in Section \ref{IntradaySDE}. We train the models on the S\&P 500 contracts that have strike prices equal or below 2,600 and test on the remaining contracts. Table \ref{Tab:summaryNewStrikePrices} reports the out-of-sample results. Figures \ref{fig:summaryNewStrikePrices1}, \ref{fig:summaryNewStrikePrices2}, and \ref{fig:summaryNewStrikePrices3} display the model-generated prices for different maturities and strike prices. In general, the neural network-SDE model is significantly more accurate than the benchmark models. 

\begin{table}[H]
    \centering
    \begin{tabular}{ |p{1.6cm}<{\centering}|p{1.4cm}<{\centering}|p{1.4cm}<{\centering}|p{1.4cm}<{\centering}|p{1.4cm}<{\centering}|p{1.4cm}<{\centering}|p{1.4cm}<{\centering}|p{1.4cm}<{\centering}|p{1.4cm}<{\centering}|}
     \hline
     \multicolumn{9}{|c|}{Experiment of out-of-sample strike prices.} \\
     \hline
     Days & \multicolumn{2}{c|}{BS Scalar}  & \multicolumn{2}{c|}{Local Volatility} & \multicolumn{2}{c|}{Heston}  & \multicolumn{2}{c|}{2D-NN} \\
     \hline
      & Call & Put  & Call & Put  & Call & Put  & Call & Put\\
     \hline
     \multirow{3}{*}{2017-09-01} & 14.929 & 9.540 & 2.763 & 16.609 & 14.663 & 9.692 & 3.601 & 3.418\\
                                & 433.947 & 179.440 & 21.242 &  645.131 & 418.509 & 187.903 & 41.155 & 24.706\\
                                & 647.346\% & 2.834\% & 100.818\% & 4.332\%  & 637.524\% & 2.865\% & 99.672\% & 1.229\%\\

    \hline
     \multirow{3}{*}{2017-10-23} & 21.278 & 11.203 & 5.846 & 12.499 & 20.596 & 11.163 & 10.010 & 11.938\\
                                & 684.422 & 190.882 & 63.031 & 359.448 & 642.026 & 188.292 & 175.755 & 233.962\\
                                & 675.996\% & 7.887\% & 145.496\% & 4.689\%  & 654.255\% & 7.744\% & 246.705\% & 7.281\%\\

    \hline
     \multirow{3}{*}{2017-11-10} & 23.299 & 11.397 & 5.451 & 14.337 & 21.147 & 10.755 & 7.855 & 9.857\\
                                & 589.513 & 201.132 & 53.887 & 516.926  & 704.976 & 179.514 & 137.897 & 223.178\\
                                & 834.499\% & 8.738\% & 226.044\% & 5.952\%  & 749.512\% & 8.053\% & 129.564\% & 4.243\%\\

    \hline
    \end{tabular}\\
    \caption{Models are trained on options with small strike prices and tested on options with large strike prices.}
    \label{Tab:summaryNewStrikePrices}
\end{table}

We observe that the 2D-NN model has the lowest combined errors (sum of the errors for call and put options) amongst all of the models. The following figures are the model-generated call and put option curves.

\subsection{Recalibration}\label{Recal}

In this section, we evaluate the neural network-SDE model over a longer time period where it is continuously recalibrated each day. The stability of the model and the benefits of recalibration are investigated. Specifically, the dataset consists of a 2 month time period (September and October of 2017), which includes 42 trading days. On each day, there are approximately 1100 contract pairs (call and put options). For each day, we randomly split the contracts into training and testing sets (80\% training and 20\% test). 

We use the two-dimensional neural network-SDE model (``2DNN") for the analysis. A continuously recalibrated model is compared with a model which is not recalibrated:
\begin{itemize}
    \item \textbf{Recalibrated Model:} On each day t, we use the model from day t-1 (the previous trading day) and train the model on the training data for day t. We test the model performance on the test dataset for day $t$ (including 2019-09-01).
        \item  \textbf{Model with No Recalibration:} We train the model on 2017-09-01 (the first day of the dataset) and test its performance on all of the data for the 42 out-of-sample trading days (including 2019-09-01).  
\end{itemize}

\begin{figure}[H]
    % \raggedleft
    \hspace{-1.5cm}\includegraphics[width=1.2\textwidth]{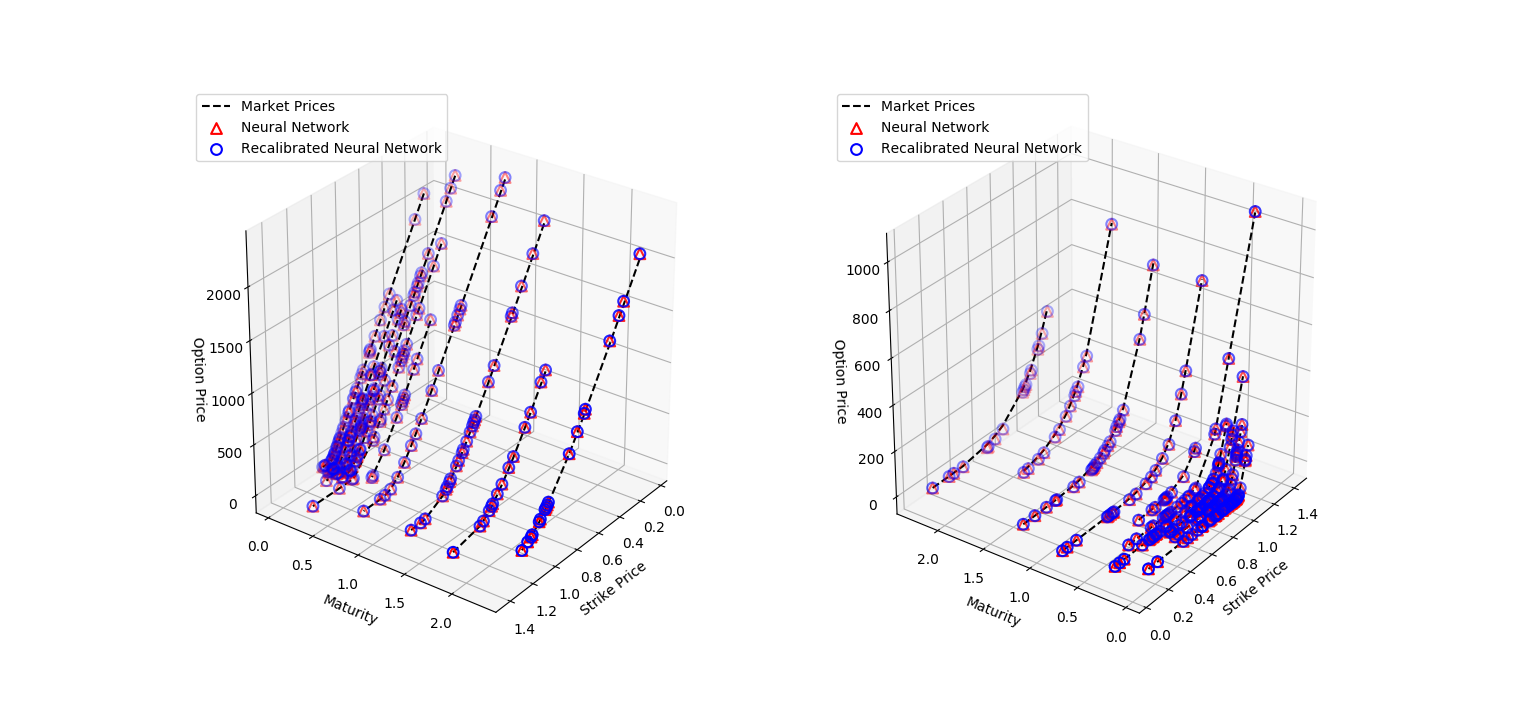}
    \caption{We compare the model performance for the out-of-sample option price curve on the 1st day (2017-09-05). The left figure is for call options and the right figure is for put options. "Market Prices" represents the true option prices. "Neural Network" is the model without recalibration while the "Recalibrated Neural Network" is the model with recalibration.}
    \label{Fig:RecalibDay1}
\end{figure}

\begin{figure}[H]
    % \raggedleft
    \hspace{-1.5cm}\includegraphics[width=1.2\textwidth]{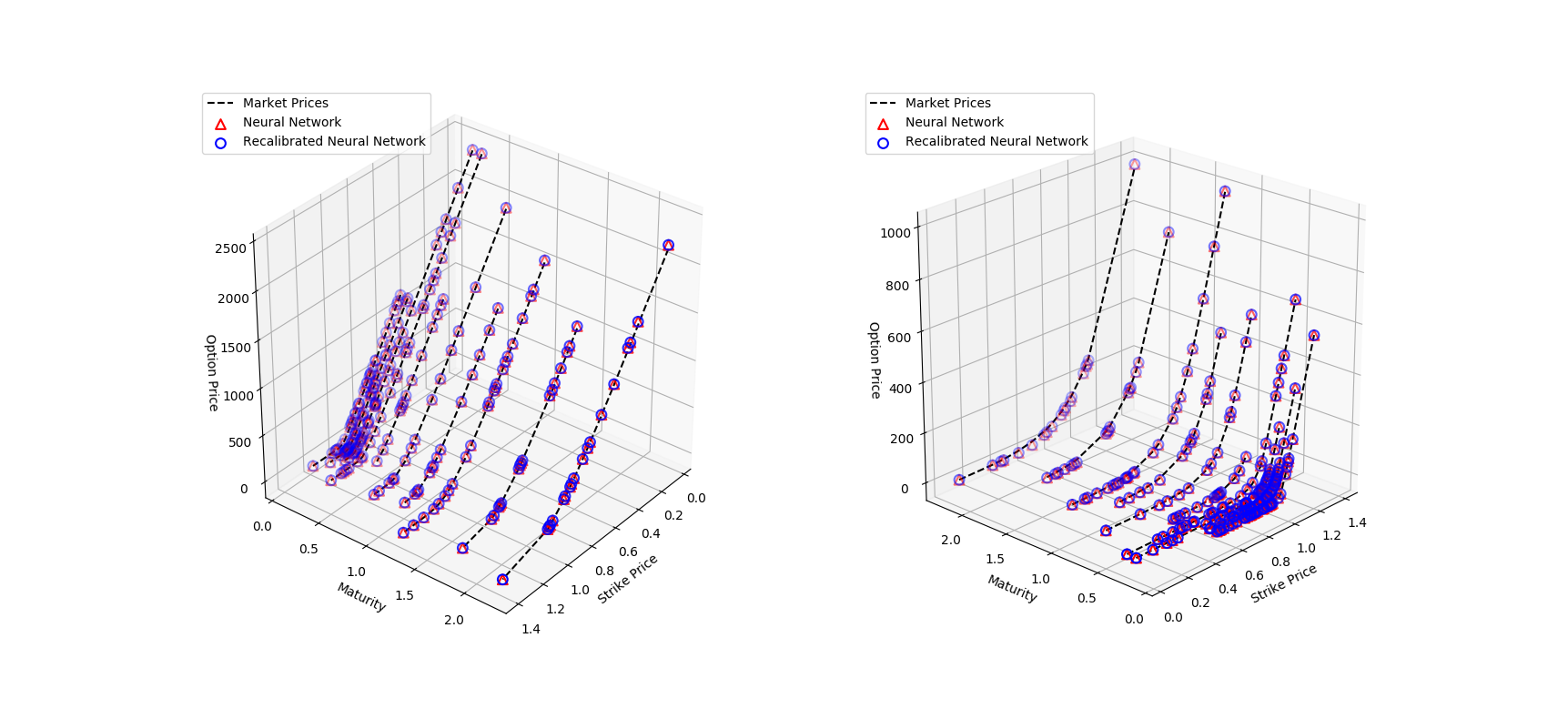}
    \caption{We compare the model performance for the the out-of-sample option price curve on the 10th day (2017-09-18). The names for the different models are the same as the previous figure \ref{Fig:RecalibDay1}.}
    % \label{fig:PartIV}
    \label{Fig:RecalibDay2}
\end{figure}

\begin{figure}[H]
    % \raggedleft
    \hspace{-1.5cm}\includegraphics[width=1.20\textwidth]{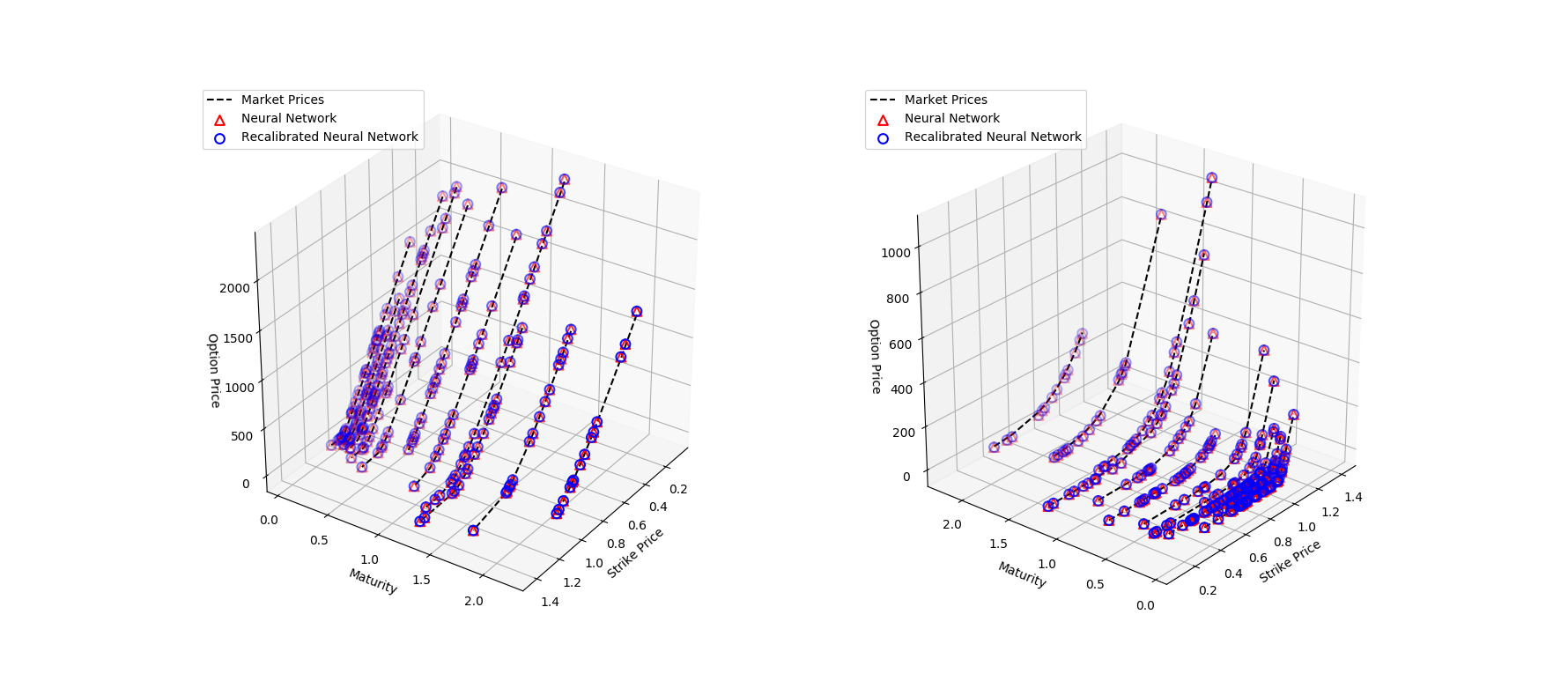}
    \caption{We compare the model performance for the out-of-sample option price curve on the 20th day (2017-10-02). The names for the different models are the same as the previous figure \ref{Fig:RecalibDay1}.}
    \label{Fig:RecalibDay3}
    % \label{fig:PartIV}
\end{figure}

\begin{figure}[H]
    % \raggedleft
    \hspace{-1.5cm}\includegraphics[width=1.20\textwidth]{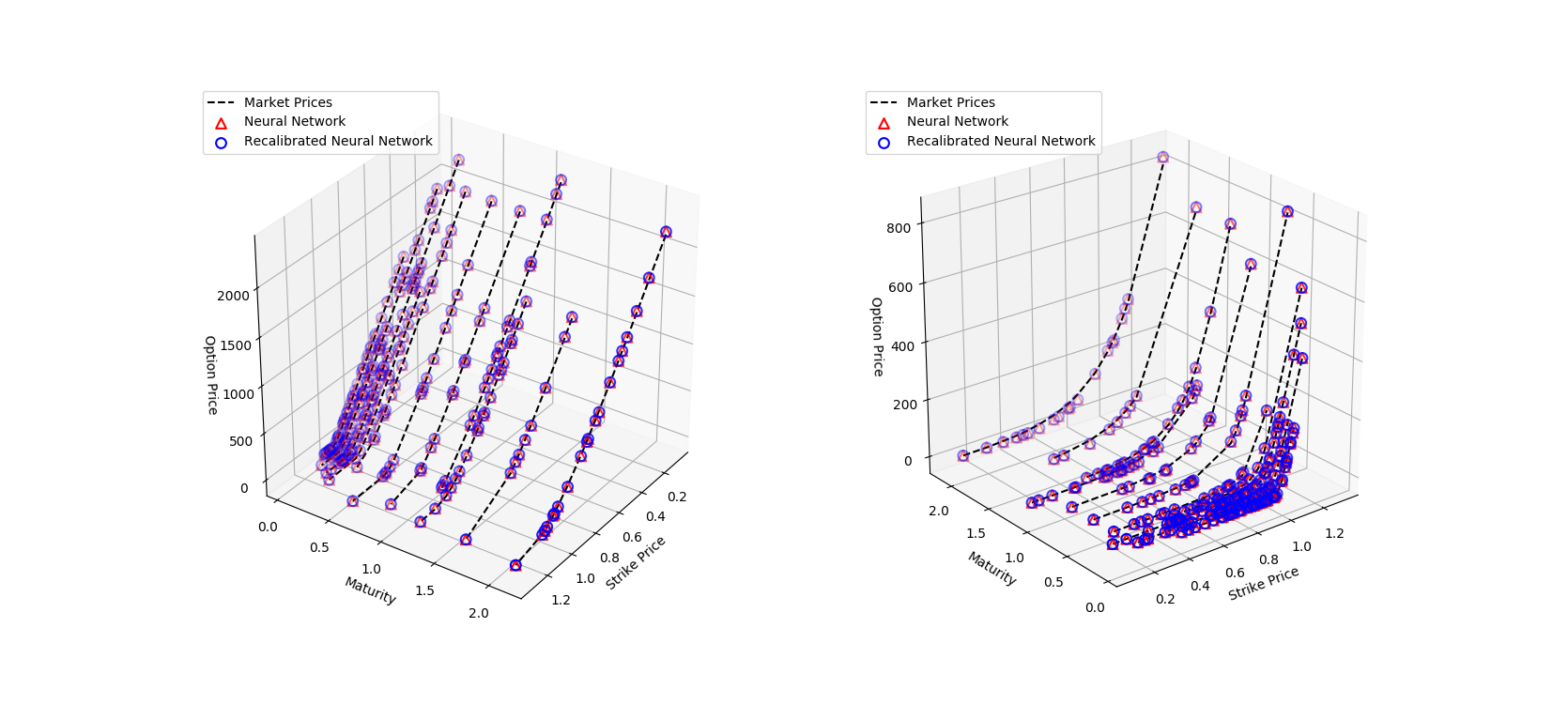}
    \caption{We compare the model performance for the out-of-sample option price curve on the 40th day (2017-10-30). The names for the different models are the same as the previous figure \ref{Fig:RecalibDay1}.}
    \label{Fig:RecalibDay4}
    % \label{fig:PartIV}
\end{figure}

Table \ref{Tab:RecalibrationTable1} compares the average performance of these two models on the test dataset. The recalibrated method outperforms the model with no recalibration, indicating that there is a benefit to recalibrating the neural network-SDE model. Figures \ref{Fig:RecalibDay1}, \ref{Fig:RecalibDay2}, \ref{Fig:RecalibDay3}, and \ref{Fig:RecalibDay4} compare the recalibrated model against the model without recalibration for days $t =1, 10, 20, 40$. The model without any recalibration performs well. Therefore, although recalibration improves accuracy, the neural network-SDE model is relatively stable across a long time period, which indicates its robustness and ability to generalize to new data. 

\begin{table}[H]
    \centering
    \begin{tabular}{ |p{4.5cm}<{\centering}|p{2.0cm}<{\centering}|p{2.0cm}<{\centering}|p{2.0cm}<{\centering}|p{2.0cm}<{\centering}| }
     \hline
     \multicolumn{5}{|c|}{Evaluation of Recalibration on Out-of-sample Data} \\
     \hline
     Days  & \multicolumn{2}{c|}{No Recalibration} & \multicolumn{2}{c|}{Recalibration} \\
     \hline
      & Call & Put & Call & Put \\
     \hline
     \multirow{3}{*}{2017-09-01 -- 2017-10-31}  & 3.684 & 2.765 &  3.230 & 2.109 \\
                                 & 34.328 & 19.197  & 32.162 & 10.243 \\
                                 & 9.439\% & 48.916\%  & 8.292\% & 47.874\% \\

     \hline
    \end{tabular}\\
    \caption{Comparison of the averaged performance of models with and without recalibration for two months. There are three rows for each cell. The first row reports the mean absolute error (MAE), the second row reports the mean-squared error (MSE), and the third row reports the relative MAE.}
    \label{Tab:RecalibrationTable1}
\end{table}

\section{Numerical Results: American Options} \label{NumericalResultsAmerican}

We will now train a neural network-SDE model on American options using the PDE optimization approach. Two datasets will be considered. The first dataset consists of American options on $100$ different stocks from the S\&P 500. On average, each stock has $\sim 530$ American option contracts. The second dataset consists of European and American options on the S\&P 100 Index.

We present several numerical experiments for the neural network-SDE models for American options: intraday out-of-sample prediction, next-day prediction, prediction for out-of-sample payoff functions, and finally training on European options and testing on American options. The first three experiments are conducted on the dataset of $100$ stocks, and the results are reported as averaged metrics across the $100$ stocks. The last experiment uses the S\&P 100 index dataset, which contains both European and American options. The neural network models, especially the 2D-NN model, have better performance than the traditional SDE models. 

Additionally, we use nonzero dividend rate for both SDE and PDE models. We sum up the dividends for the past 1 year and divide them by the closing price for single stocks, while we use 1.91\% for S\&P $100$ index options.

\subsection{Intraday Out-of-sample Prediction} \label{AmericanIntraday}

\begin{figure}[H]
    % \raggedleft
    \hspace{-1.5cm}
    \hspace{1.0cm}\includegraphics[width=1.03\textwidth]{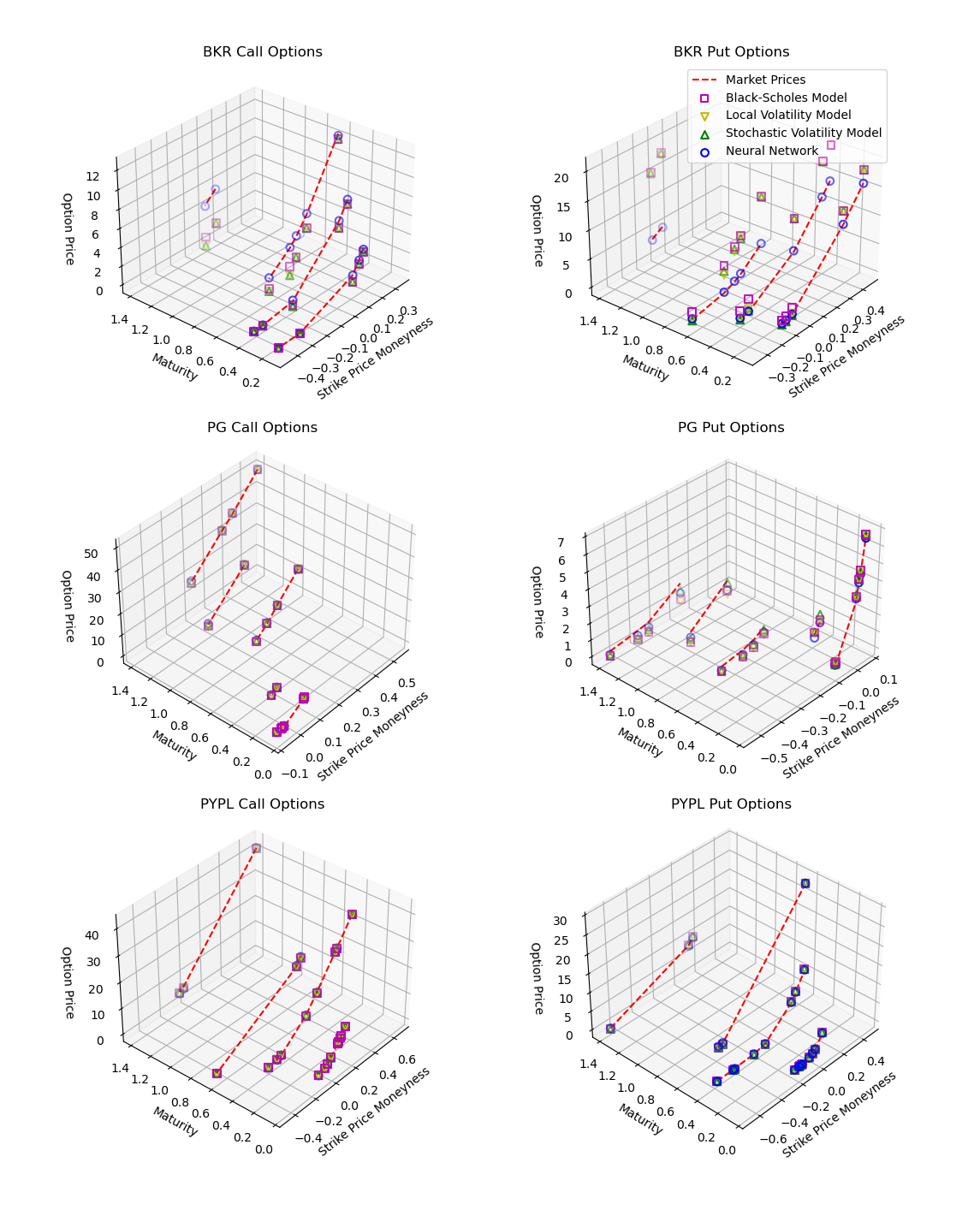}
    \vspace{-2em}
    \caption{The intraday out-of-sample option curve for three stocks: "BKR", "PG" and "PYPL". The left and right figures are call and put options, respectively. The date is 2017-09-01. "Market Prices" represents the true option prices. "Black-Scholes Model"  refers to the "BS-Scalar" model. "Local Volatility Model" refers to the "NNLV" model. 
    "Neural Network" refers to the "2D-NN" model, while the "Stochastic Volatility Model" refers to the Heston model.}
    \label{Fig:intra_BKR}
\end{figure}

The intraday pricing accuracy for the neural network-SDE model is reported in Table \ref{Tab:American1}. The results are averaged across the contracts from $100$ different stocks. For each stock (on the date 2017-09-01), we select 80\% of the American call and put options for training and the remaining 20\% of the contracts for testing. The call and put options are selected as pairs (i.e., with the same strike and maturity) so that the out-of-sample options will be at different strikes or maturities. Table \ref{Tab:American2} compares different neural network-SDE models. Figure \ref{Fig:intra_BKR} compares the model-generated prices for different maturities and strike prices for several stocks. 

\begin{table}[H]
    \centering
    \begin{tabular}{ |p{2.2cm}<{\centering}|p{1.9cm}<{\centering}|p{1.9cm}<{\centering}|p{1.9cm}<{\centering}|p{1.9cm}<{\centering}|p{1.9cm}<{\centering}|p{1.9cm}<{\centering}|}
     \hline
     \multicolumn{7}{|c|}{Intraday Prediction for Different Models.} \\
     \hline
      & \multicolumn{2}{c|}{BS Scalar}  & \multicolumn{2}{c|}{Heston} & \multicolumn{2}{c|}{2D-NN}\\
     \hline
      & Call & Put & Call & Put & Call & Put \\
     \hline
     \multirow{3}{*}{2017-09-01}
      & 0.433 & 0.448 &  0.401 & 0.416 & 0.345 & 0.352\\
      & 1.369 & 1.315 &  1.166 & 1.188 & 0.857 & 0.976\\
      & 40.063\% & 38.253\%  & 33.623\% & 34.952\% & 32.188\% & 28.121\%\\
     \hline
    \end{tabular}\\
    \vspace{2em}
    \caption{Intraday prediction for different SDE models. There are three rows for each column. The first row reports MAE, the second row reports MSE, and the third row reports relative MAE.}
    \label{Tab:American1}
\end{table}
\vspace{-2em}

\begin{table}[H]
    \centering
    \begin{tabular}{ |p{2.2cm}<{\centering}|p{1.9cm}<{\centering}|p{1.9cm}<{\centering}|p{1.9cm}<{\centering}|p{1.9cm}<{\centering}|p{1.9cm}<{\centering}|p{1.9cm}<{\centering}|}
     \hline
     \multicolumn{7}{|c|}{Intraday Prediction for Deep Learning Models.} \\
     \hline
     2017-09-01 & \multicolumn{2}{c|}{NNLV} & \multicolumn{2}{c|}{2D-NN-Heston} & \multicolumn{2}{c|}{2D-NN} \\
     \hline
      & Call & Put & Call & Put & Call & Put \\
     \hline
     \multirow{3}{*}{2017-09-01}
      & 0.423 & 0.424 & 0.358 & 0.340 & 0.345 & 0.352\\
       & 1.220 & 1.127 & 1.026 & 1.084 & 0.857 & 0.976\\
      & 39.194\% & 37.005\% & 29.633\% & 30.148\% & 32.188\% & 28.121\%\\

     \hline
    \end{tabular}\\
    \vspace{2em}
    \caption{Intraday prediction for deep learning SDE models. There are three rows for each column. The first row reports MAE, the second row reports MSE, and the third row reports relative MAE.}
    \label{Tab:American2}
\end{table}

\subsection{Next-day Out-of-sample Performance}\label{sectionPDENextDay}

\begin{figure}[H]
    % \raggedleft
    \hspace{-1.5cm}
    \hspace{1.0cm}\includegraphics[width=1.03\textwidth]{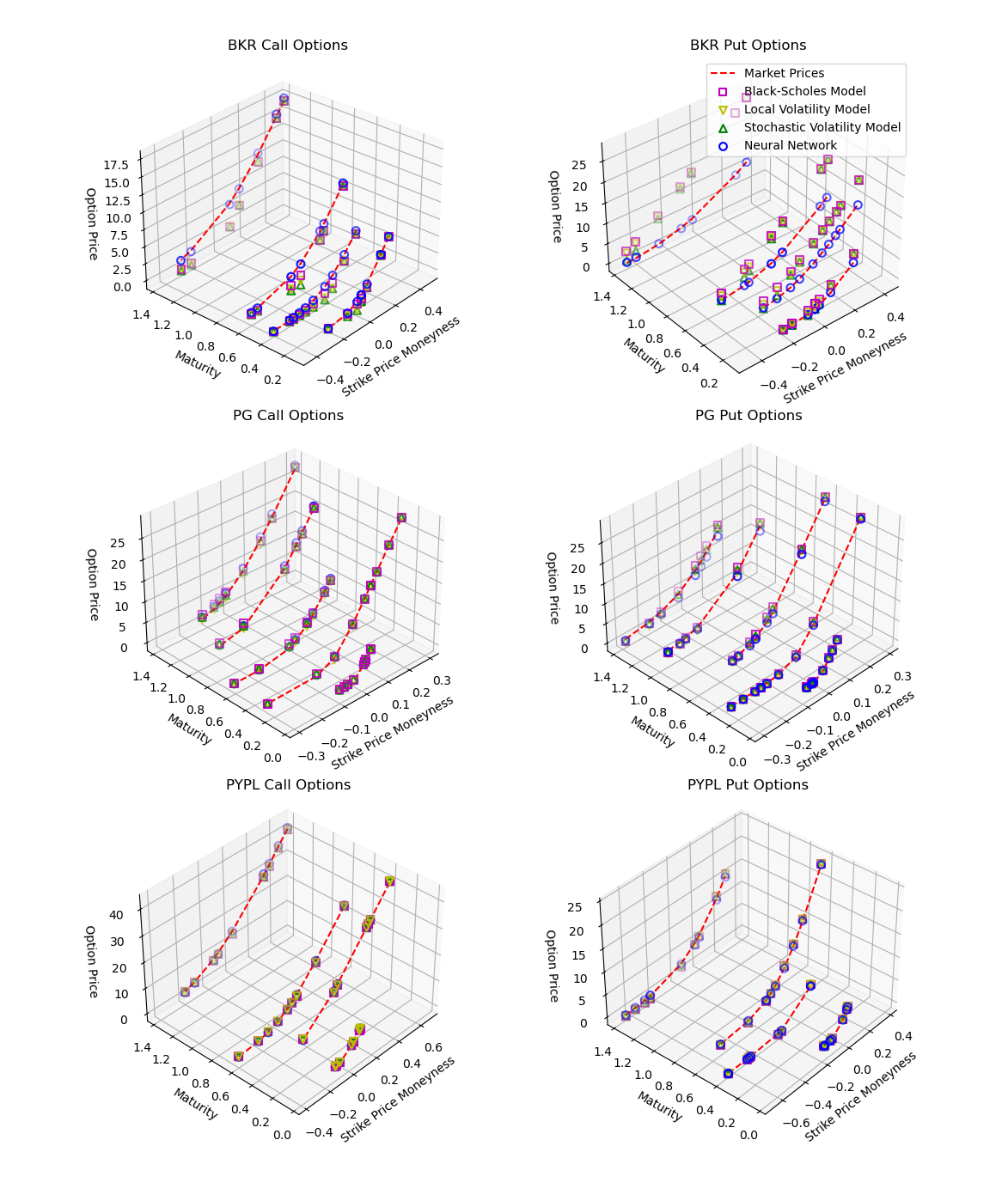}
    \vspace{-2em}
    \caption{The next day out-of-sample option price curve for three stocks: "BKR", "PG" and "PYPL". The left and right figures are call and put options, respectively. It contains randomly selected data points from 2017-09-01 for clear visualization.   "Market Prices" represents the true option prices. "Black-Scholes Model"  refers to the "BS-Scalar" model. "Local Volatility Model" refers to the "NNLV" model. 
    "Neural Network" refers to the "2D-NN", while the "Stochastic Volatility Model" refers to the Heston model.}
    \label{Fig:nextday}
\end{figure}

In this experiment, we train the models on day $t$ and evaluate their accuracy for pricing contracts on day $t+1$. For $100$ different stocks, the models are trained on American options data from a single data 2017-09-01 and then evaluated on the next trading day 2017-09-05. Table \ref{Tab:American3} compares the neural network-SDE with the several benchmark SDE models. There are two Black-Scholes models in the table: BS-scalar and BS-LV. The former model calibrates the implied volatility parameter as the same constant scalar for all the contracts in a day, while the latter model calibrates a different implied volatility parameter for each call and put option pair. Table \ref{Tab:American4} compares different variations of the neural network-SDE model. The performance of these models is visualized in Figure \ref{Fig:nextday} for several stocks, where the model-generated prices are plotted for different maturities and strike prices. The neural-network SDE model outperforms the benchmark SDE models for longer maturities.

\begin{table}[H]
    \centering
    \begin{tabular}{ |p{2.2cm}<{\centering}|p{1.3cm}<{\centering}|p{1.3cm}<{\centering}|p{1.3cm}<{\centering}|p{1.3cm}<{\centering}|p{1.3cm}<{\centering}|p{1.3cm}<{\centering}|p{1.3cm}<{\centering}|p{1.3cm}<{\centering}|p{1.3cm}<{\centering}|}
     \hline
     \multicolumn{9}{|c|}{Experiment of the Next Day Prediction for Different Models.} \\
     \hline
      & \multicolumn{2}{c|}{BS Scalar} & \multicolumn{2}{c|}{BS-LV} & \multicolumn{2}{c|}{Heston} & \multicolumn{2}{c|}{2D-NN}\\
     \hline
      & Call & Put & Call & Put  & Call & Put & Call & Put\\
     \hline
     \multirow{3}{*}{2017-09-05}
      & 0.400 & 0.433 & 0.307 & 0.314 & 0.372 & 0.391 & 0.264 & 0.286\\
      & 1.349 & 1.237 & 0.487 & 0.747 & 1.136 & 1.106 & 0.552 & 0.446\\
      & 38.651\% & 36.421\%  & 36.757\%  & 36.694\% & 31.416\% & 33.596\% & 27.597\% & 25.475\%\\
     \hline

    \end{tabular}\\
    \caption{Accuracy for the next-day prediction for different SDE models. There are three rows for each column. The first row reports MAE, the second row reports MSE, and the third row reports relative MAE. }
    \label{Tab:American3}
\end{table}

\vspace{-2.em}
\begin{table}[H]
    \centering
    \begin{tabular}{ |p{2.5cm}<{\centering}|p{1.8cm}<{\centering}|p{1.8cm}<{\centering}|p{1.8cm}<{\centering}|p{1.8cm}<{\centering}|p{1.8cm}<{\centering}|p{1.8cm}<{\centering}|}
     \hline
     \multicolumn{7}{|c|}{Experiment of the Next Day Prediction for Deep Learning Models.} \\
     \hline
      & \multicolumn{2}{c|}{NNLV} & \multicolumn{2}{c|}{2D-NN-Heston} & \multicolumn{2}{c|}{2D-NN}\\
     \hline
      & Call & Put & Call & Put & Call & Put \\
     \hline
     \multirow{3}{*}{2017-09-05}
      & 0.387 & 0.419 & 0.352 & 0.327 & 0.264 & 0.286 \\
      &1.136 & 1.156 & 1.253 & 0.704 & 0.552 & 0.446 \\
      & 35.249\% & 35.586\% & 27.778\% & 29.756\%  & 27.597\% & 25.475\% \\
     \hline

    \end{tabular}\\
    \caption{Accuracy of the next-day prediction for deep learning SDE models. There are three rows for each column. The first row reports MAE, the second row reports MSE, and the third row reports relative MAE.}
    \label{Tab:American4}
\end{table}
\vspace{-2.em}

\subsection{Out-of-sample Payoff Functions} \label{AmericanOutofSamplePayoffFunctions}

\begin{figure}[H]
    % \raggedleft
    \hspace{-1.5cm}
    \hspace{1.0cm}\includegraphics[width=1.05\textwidth]{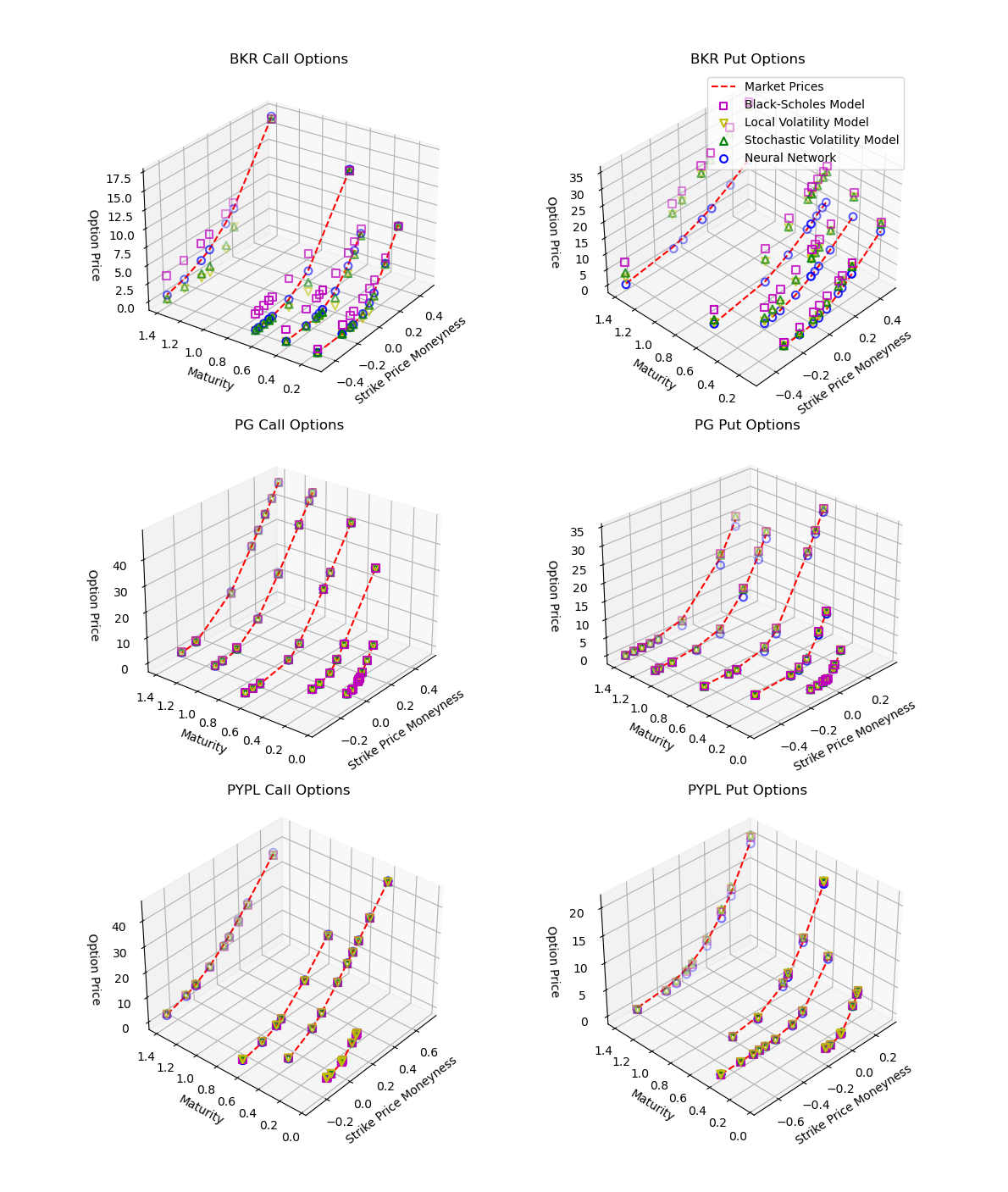}
    \vspace{-2em}
    \caption{The out-of-sample payoff function results for 3 stocks "BKR", "PG" and "PYPL". The left and right figures are call and put options, respectively. The plot consists of randomly selected data points (contracts) from 2017-09-01 for clear visualization of the results. "Market Prices" rrefers to the true option prices. "Black-Scholes Model"  refers to the "BS-Scalar" model. "Local Volatility Model" refers to the "NNLV" model. 
    "Neural Network" refers to the "2D-NN", while the "Stochastic Volatility Model" refers to the Heston model.}
    \label{Fig:OutofsamplePDE}
\end{figure}

In this experiment across $100$ stocks on the date 2017-09-01, we will train on American call options and then evaluate the model performance on American put options. Table \ref{Tab:intradayPDE3} reports the results (averaged across the $100$ stocks) comparing the neural network-SDE model with the benchmark models. Table \ref{Tab:intradayPDE4} compares different variations of the neural network-SDE model. The performance of the models is visualized in Figure \ref{Fig:OutofsamplePDE}. Similar to the results in Section \ref{sectionPDENextDay}, the neural network-SDE model is more accurate for longer maturities.

\begin{table}[H]
    \centering
    \begin{tabular}{ |p{2.2cm}<{\centering}|p{1.3cm}<{\centering}|p{1.3cm}<{\centering}|p{1.3cm}<{\centering}|p{1.3cm}<{\centering}|p{1.3cm}<{\centering}|p{1.3cm}<{\centering}|p{1.3cm}<{\centering}|p{1.3cm}<{\centering}|p{1.3cm}<{\centering}|}
     \hline
     \multicolumn{9}{|c|}{Experiment of Price Prediction for Different Models.} \\
     \hline
      & \multicolumn{2}{c|}{BS Scalar} & \multicolumn{2}{c|}{BS-LV} & \multicolumn{2}{c|}{Heston} & \multicolumn{2}{c|}{2D-NN}\\
     \hline
      & Call & Put & Call & Put  & Call & Put & Call & Put\\
     \hline
     \multirow{3}{*}{2017-09-01}
      & 0.450 & 0.531 &  0.258 & 0.415 & 0.412 & 0.492 & 0.269 & 0.464\\
      & 1.273 & 2.018 &  0.355 & 1.337 & 1.142 & 1.745 & 0.693 & 1.158\\
      & 53.433\% & 45.518\%  & 36.061\% & 45.653\% & 35.262\% & 36.340\% & 26.761\% & 38.363\%\\
     \hline
    \end{tabular}\\
    \caption{Accuracy for the options price prediction for different models. There are three rows for each column. The first row reports MAE, the second row reports MSE, and the third row reports relative MAE. }
    \label{Tab:intradayPDE3}
\end{table}

\vspace{-1.5em}
\begin{table}[H]
    \centering
    \begin{tabular}{ |p{2.5cm}<{\centering}|p{1.8cm}<{\centering}|p{1.8cm}<{\centering}|p{1.8cm}<{\centering}|p{1.8cm}<{\centering}|p{1.8cm}<{\centering}|p{1.8cm}<{\centering}|}
     \hline
     \multicolumn{7}{|c|}{Experiment of Price Prediction for Deep Learning Models.} \\
     \hline
     & \multicolumn{2}{c|}{NNLV} & \multicolumn{2}{c|}{2D-NN-Heston} & \multicolumn{2}{c|}{2D-NN}\\
     \hline
     & Call & Put & Call & Put & Call & Put \\
     \hline
     \multirow{3}{*}{2017-09-01}
      & 0.430 & 0.487 & 0.361 & 0.420 & 0.269 & 0.464 \\
      & 1.196 & 1.528 &1.097 & 1.391 & 0.693 & 1.158 \\
      & 44.740\% & 38.576\% & 31.903\% & 36.347\%  & 26.761\% & 38.363\%  \\
     \hline
    \end{tabular}\\
    \caption{Accuracy of the options price prediction for deep learning SDE models.There are three rows for each column. The first row reports MAE, the second row reports MSE, and the third row reports relative MAE.}
    \label{Tab:intradayPDE4}
\end{table}

\vspace{-2em}

% \vspace{-2em}

\vspace{-1em}

% \vspace{-1em}
\subsection{Training on European Options and Testing on American Options} \label{SP100EuropeanAmerican}

In this section, we train the models on European option data for the S\&P $100$ index. Then, we evaluate the model performance for pricing American options on the S\&P $100$ index. This an example of training a model on a certain type of payoff function and then evaluating whether it can generalize to accurately price a financial derivative with a different payoff function. The results are reported in Tables \ref{Tab1:intradaySP100} and \ref{Tab2:intradaySP100}. Figure \label{Fig1:SP100} provides a visualization of the model accuracy for different maturities and strike prices. Figure \ref{Fig2:SP100} plots the model-generated prices versus the strike price for a specific maturity.

\begin{table}[H]
    \centering
    \begin{tabular}{ |p{2.2cm}<{\centering}|p{1.3cm}<{\centering}|p{1.3cm}<{\centering}|p{1.3cm}<{\centering}|p{1.3cm}<{\centering}|p{1.3cm}<{\centering}|p{1.3cm}<{\centering}|p{1.3cm}<{\centering}|p{1.3cm}<{\centering}|p{1.3cm}<{\centering}|}
     \hline
     \multicolumn{9}{|c|}{Experiments of Different Models for S\&P 100 option.} \\
     \hline
      & \multicolumn{2}{c|}{BS Scalar} & \multicolumn{2}{c|}{BS-LV} & \multicolumn{2}{c|}{Heston} & \multicolumn{2}{c|}{2D-NN}\\
     \hline
      & Call & Put & Call & Put  & Call & Put & Call & Put\\
     \hline
     \multirow{3}{*}{2017-09-01}
      & 2.062 & 1.773 & 1.893  & 1.555 & 2.132 & 1.789 & 2.030 & 1.336\\
      & 13.141 & 10.835 &  10.020 & 7.706 & 10.542 & 9.200 & 8.064 & 4.665\\
      & 42.600\% & 65.815\%  & 49.956\% & 75.016\% &  63.313\% & 67.480\% & 35.079\% & 65.195\%\\
     \hline
    \end{tabular}\\
    \caption{Accuracy of PDE models trained on European options and evaluated out-of-sample on American options. There are three rows for each column. The first row reports MAE, the second row reports MSE, and the third row reports relative MAE. }
    \label{Tab1:intradaySP100}
\end{table}

\vspace{-2em}
\begin{table}[H]
    \centering
    \begin{tabular}{ |p{2.5cm}<{\centering}|p{1.8cm}<{\centering}|p{1.8cm}<{\centering}|p{1.8cm}<{\centering}|p{1.8cm}<{\centering}|p{1.8cm}<{\centering}|p{1.8cm}<{\centering}|}
     \hline
     \multicolumn{7}{|c|}{Experiments of Deep Learning Models for S\&P 100 option.} \\
     \hline
     & \multicolumn{2}{c|}{NNLV} & \multicolumn{2}{c|}{2D-NN-Heston} & \multicolumn{2}{c|}{2D-NN}\\
     \hline
     & Call & Put & Call & Put & Call & Put \\
     \hline
     \multirow{3}{*}{2017-09-01}
      & 1.963 & 1.653 & 1.904 & 1.649 & 2.030 & 1.336 \\
      & 10.095 & 8.723 & 8.339 & 9.951 & 8.064 & 4.665 \\
      & 42.910\% & 63.484\% & 45.455\% & 70.487\%  & 35.079\% & 65.195\%  \\
     \hline
    \end{tabular}\\
    \caption{Results for PDE models trained on European options and evaluated out-of-sample on American options. There are three rows for each column. The first row reports MAE, the second row reports MSE, and the third row reports relative MAE.}
    \label{Tab2:intradaySP100}
\end{table}

\vspace{-1em}
\begin{figure}[H]
    % \raggedleft
    \hspace{-1cm}\includegraphics[width=1.1\textwidth]{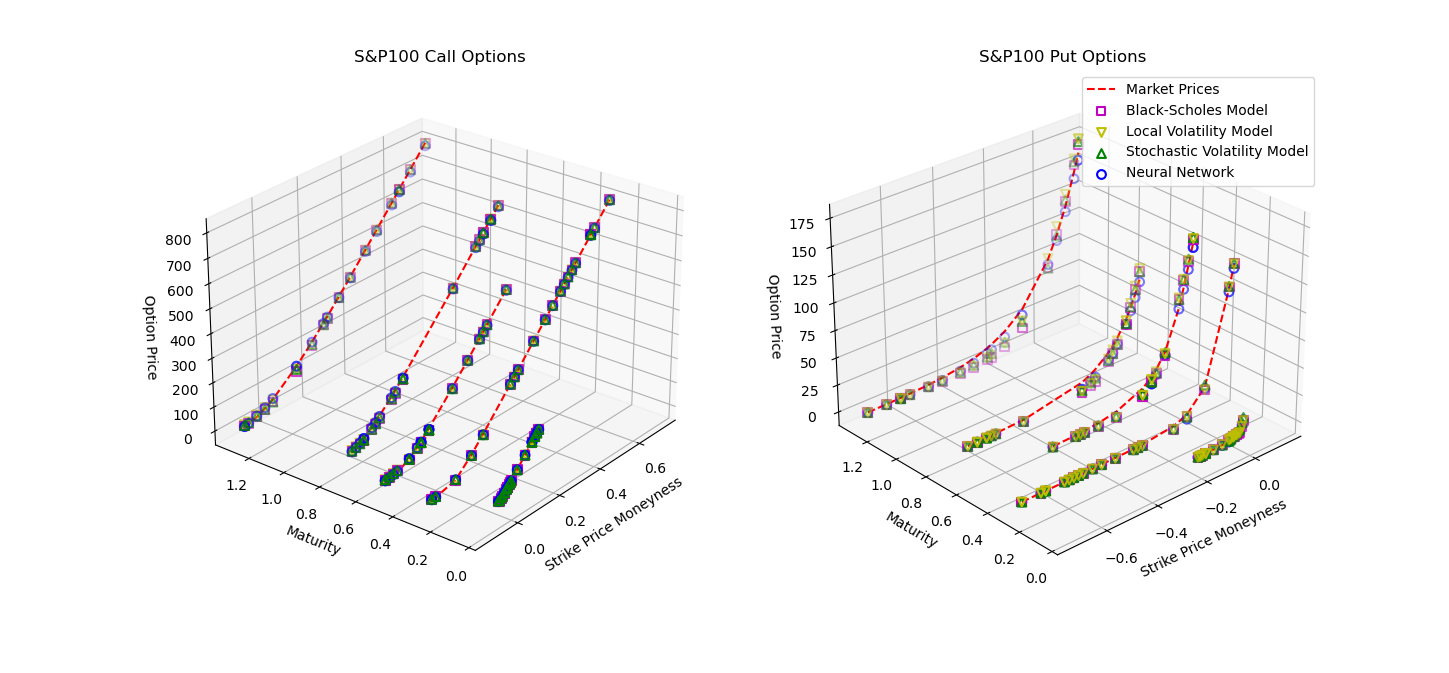}
    \caption{Numerical results for the out-of-sample payoff function for the S\&P 100 index option. The left and right figures are call and put options, respectively. Randomly selected data points from 2017-09-01 are plotted for clear visualization.  "Market Prices" refers to the true option prices. "Black-Scholes Model"  refers to the "BS-Scalar" model. "Local Volatility Model" refers to the "NNLV" model. 
    "Neural Network" refers to the "2D-NN", while the "Stochastic Volatility Model" refers to the Heston model.}
        \label{Fig1:SP100}
\end{figure}

\vspace{-1em}
\begin{figure}[H]
    % \raggedleft
    \hspace{-1cm}\includegraphics[width=1.1\textwidth]{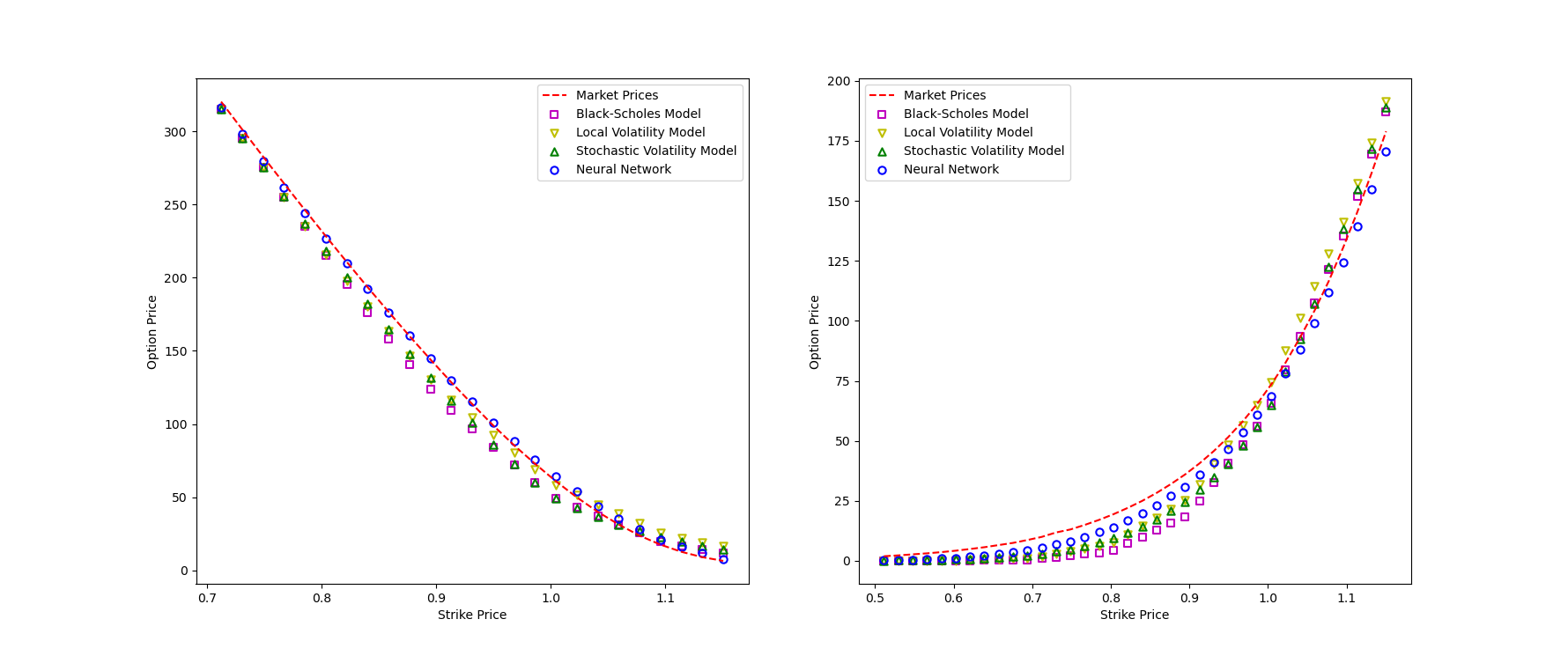}
    \caption{Numerical results for the out-of-sample payoff function for the S\&P 100 index option when the maturity is 476 days. The left and right figures are call and put options, respectively. The date is 2017-09-01. "Market Prices" refers to the true option prices. "Black-Scholes Model"  refers to the "BS-Scalar" model. "Local Volatility Model" refers to the "NNLV" model. 
    "Neural Network" refers to the "2D-NN", while the "Stochastic Volatility Model" refers to the Heston model.}
     \label{Fig2:SP100}
\end{figure}

\section{Numerical Results: Hedging} \label{Hedging}
Once the neural-network SDE models are trained, they can also be used to estimate delta hedges for options. In this section, we evaluate the performance of the neural-network SDE models for hedging options. 

\subsection{Next Day Hedging}
This section evaluates the hedging performance for the next day out-of-sample experiment from Section \ref{NextDaySDE}. The results for the different models are described below.

\begin{table}[H]
    \centering
    \begin{tabular}{ |p{1.6cm}<{\centering}|p{1.4cm}<{\centering}|p{1.4cm}<{\centering}|p{1.4cm}<{\centering}|p{1.4cm}<{\centering}|p{1.4cm}<{\centering}|p{1.4cm}<{\centering}|p{1.4cm}<{\centering}|p{1.4cm}<{\centering}| }
     \hline
     \multicolumn{9}{|c|}{Next day delta hedging performance for traditional SDE models} \\
     \hline
     Days &  \multicolumn{2}{c|}{BS} & \multicolumn{2}{c|}{Local Volatility} & \multicolumn{2}{c|}{Heston}  & \multicolumn{2}{c|}{2D-NN} \\
     \hline
      & Call & Put & Call & Put & Call & Put & Call & Put\\
     \hline
     \multirow{3}{*}{2017-09-01} & 0.971 & 0.777  & 1.466 & 1.226 & 1.255 & 0.135 & 1.047 & 0.988\\
                                & 1.653 & 1.183  & 3.374 & 2.872 & 2.525 & 2.403 & 2.500 & 2.412\\
                                & 5.533\% &  8.247\% & 23.483\% & 17.376\% &  17.940\% & 16.903\% & 4.233\% & 14.556\%\\
     \hline
     \multirow{3}{*}{2017-10-23} & 0.783 & 0.291  & 0.832 & 0.247 &  0.827 & 0.232 & 0.779 & 0.264\\
                                & 0.746  & 0.177  & 0.841 & 0.162 & 0.814 & 0.153 & 0.745 & 0.159\\
                                &  3.104\% &  5.059\% & 4.404\% & 3.479\% &  3.938\% & 3.312\% & 2.565\% & 3.778\%\\
     \hline
     \multirow{3}{*}{2017-11-10}& 0.249 & 0.218  & 0.261 & 0.248 & 0.273 & 0.261 & 0.260 & 0.241\\
                                & 0.098  & 0.108  & 0.107 & 0.129 & 0.112 & 0.141 & 0.116 & 0.139\\
                                & 2.080\% &  6.291\% & 3.046\% & 7.020\% &  3.649\% & 7.059\% & 1.583\% & 6.840\%\\
     \hline
     \multirow{3}{*}{2017-12-07} &1.478 & 0.233  & 1.571 & 0.536 &  1.755 & 0.581 & 1.663 & 0.672\\
                                & 2.570  & 0.145  & 3.027 & 0.550 & 3.514 & 0.615  & 3.436 & 0.961\\
                                & 4.506\% &  3.497\% & 11.620\% & 8.893\% &  10.174\% & 9.334\% & 3.346\% & 7.376\%\\
     \hline
     \multirow{3}{*}{2018-01-26} & 5.756 & 0.977  & 5.753 & 1.441 &  5.901  & 1.453 & 5.616 & 1.483\\
                                & 37.351 & 1.936  & 36.694 & 3.922 &  38.410 & 3.975 & 35.617 & 4.423\\
                                & 7.447\% &  13.561\% & 9.286\% & 20.016\% &  9.292\% & 20.722\% & 5.857\% & 18.378\%\\
     \hline
    \end{tabular}\\
    \caption{Next-day delta hedging performance for different models. The three rows in each cell are MAE, MSE, relative MAE, respectively.}
    \label{Tab:ATMHedgingsummary1}
\end{table}

\begin{table}[H]
    \centering
    \begin{tabular}{ |p{1.6cm}<{\centering}|p{1.4cm}<{\centering}|p{1.4cm}<{\centering}|p{1.4cm}<{\centering}|p{1.4cm}<{\centering}|p{1.4cm}<{\centering}|p{1.4cm}<{\centering}|p{1.4cm}<{\centering}|p{1.4cm}<{\centering}| }
     \hline
     \multicolumn{9}{|c|}{Next day delta hedging for neural network-SDE models} \\
     \hline
     Days & \multicolumn{2}{c|}{NNLV} &  \multicolumn{2}{c|}{SDENN} & \multicolumn{2}{c|}{SDENN-Drift} & \multicolumn{2}{c|}{2D-NN} \\
     \hline
      & Call & Put & Call & Put & Call & Put & Call & Put\\
     \hline
     \multirow{3}{*}{2017-09-01} & 1.293 & 1.149 & 1.302 & 1.160 & 1.194 & 1.134 & 1.047 & 0.988 \\
                                & 3.362  & 3.058 & 3.398 & 3.100 &  3.047 &  2.973 & 2.500 & 2.412 \\
                                &  8.376\% & 15.603\% & 9.318\% & 15.605\% &  9.275\% &  14.907\% & 4.233\% & 14.556\% \\
     \hline
     \multirow{3}{*}{2017-10-23} &  0.831 & 0.248 & 0.830 & 0.249 & 0.799 & 0.253  & 0.779 & 0.264  \\
                                & 0.845  & 0.160 & 0.843 & 0.161 &  0.780 & 0.157  & 0.745 & 0.159 \\
                                &  2.895\% & 3.415\% & 3.063\% & 3.478\% &  3.016\% &  3.529\% & 2.565\% & 3.778\% \\
     \hline
     \multirow{3}{*}{2017-11-10}& 0.268  & 0.263 & 0.264 & 0.263 & 0.260 & 0.257  & 0.260 & 0.241 \\
                                & 0.133  & 0.168 & 0.131 & 0.168 &  0.125 & 0.159  & 0.116 & 0.139 \\
                                &  1.457\% & 7.011\% & 1.569\% & 6.999\% &  1.777\% &  6.889\% & 1.583\% & 6.840\% \\
     \hline
     \multirow{3}{*}{2017-12-07} &  1.804 & 0.817 & 1.815 & 0.780 & 1.687 & 0.745  & 1.663 & 0.672 \\
                                & 4.180  & 1.243 & 4.134 & 1.146 & 3.618 & 1.019  & 3.436 & 0.961 \\
                                &  2.582\% & 9.853\% & 3.416\% & 9.631\% &  2.441\% &  9.416\% & 3.346\% & 7.376\% \\
     \hline
     \multirow{3}{*}{2018-01-26} &  5.962 & 1.696 & 5.964 & 1.681 & 5.780 & 1.640  & 5.616 & 1.483 \\
                                &  40.480 & 5.394 & 40.411 & 5.290 &  37.995 & 5.048  & 35.617 & 4.423 \\
                                &  7.041\% & 20.461\% & 7.401\% & 20.402\% &  6.788\% &  20.176\% & 5.857\% & 18.378\% \\
     \hline
    \end{tabular}\\
    \caption{Next-day delta hedging performance for the neural network SDE models. The three rows in each cell are MAE, MSE, relative MAE, respectively.}
    \label{Tab:ATMHedgingsummary2}
\end{table}
\vspace{-0.5cm}
We observe that the neural-network-based SDE model outperforms Dupire's local volatility model and the Heston model while slightly underperforming the Black-Scholes model.

\subsection{At-The-Money Hedging} 
In this section, we consider a single time horizon $T=1$ (one-day delta hedging). Furthermore, we select the contracts that satisfy the conditions below:
\begin{itemize}
    \item The maturity is between one month and one year.
    \item Contracts where the strike price is between \$2200 and \$2800.
\end{itemize}
\vspace{-0.15cm}
The at-the-money hedging results for the different models are presented below. 

\begin{table}[H]
    \centering
    \begin{tabular}{ |p{1.6cm}<{\centering}|p{1.4cm}<{\centering}|p{1.4cm}<{\centering}|p{1.4cm}<{\centering}|p{1.4cm}<{\centering}|p{1.4cm}<{\centering}|p{1.4cm}<{\centering}|p{1.4cm}<{\centering}|p{1.4cm}<{\centering}| }
     \hline
     \multicolumn{9}{|c|}{Next day at-the-money delta hedging for traditional SDE models} \\
     \hline
     Days &  \multicolumn{2}{c|}{BS} & \multicolumn{2}{c|}{Local Volatility} & \multicolumn{2}{c|}{Heston} & \multicolumn{2}{c|}{2D-NN} \\
     \hline
      & Call & Put & Call & Put & Call & Put & Call & Put\\
     \hline
     \multirow{3}{*}{2017-09-01}  & 1.703 & 1.528 & 2.671 & 2.533  & 2.358 & 2.274 & 2.482 & 2.458  \\
                                 & 3.541 & 2.519 & 7.650 & 7.129  & 5.826 & 5.517 & 7.645 & 7.317\\
                                 & 2.922\% & 6.471\% & 10.040\% &  12.468\% & 9.563\% &  12.393\% & 5.538\% & 11.443\%\\
     \hline
     \multirow{3}{*}{2017-10-23} & 0.683 & 0.497 & 0.815 & 0.413  & 0.782 & 0.395 & 0.748 & 0.432\\
                                 & 0.627 & 0.352 & 0.876 & 0.291  & 0.785 & 0.280 & 0.757 & 0.269\\
                                & 4.512\% & 3.049\% & 7.089\% &  1.790\% & 6.085\% &  1.568\% & 3.853\% & 2.003\%\\
     \hline
     \multirow{3}{*}{2017-11-10}  & 0.328 & 0.373 & 0.338 & 0.450  & 0.373 & 0.463 & 0.347 & 0.455\\
                                 & 0.163 & 0.212 & 0.173 & 0.278  & 0.191 & 0.287 & 0.202 & 0.309\\
                                 & 2.676\% & 2.649\% &  4.101\% &  3.558\% & 5.081\% &  3.592\% & 2.293\% & 3.403\%\\
     \hline
     \multirow{3}{*}{2017-12-07}  & 1.652 & 0.263 &  1.926 & 0.774  & 2.136 & 0.862 & 2.260 & 1.071\\
                                 & 3.003 & 0.109 & 4.135 & 0.823  & 4.869 & 0.976 & 5.584 & 1.713\\
                                 & 2.669\% & 1.376\% &  4.716\% &  7.871\% & 5.020\% &  8.602\% & 4.344\% & 6.392\%\\
     \hline
     \multirow{3}{*}{2017-01-26}  & 7.708 & 1.477 & 7.391 & 2.247  & 7.789 & 2.543 & 7.358 & 2.237\\
                                 & 59.894 & 2.469 & 55.409 & 5.760  & 61.274 & 7.051 & 54.847 & 5.764\\
                                 & 2.849\% & 10.636\% & 2.811\% &  18.075\% & 2.868\% &  19.566\% & 2.793\% & 15.803\%\\
     \hline
    \end{tabular}\\
    \caption{Next-day at-the-money delta hedging performance for the different models. The three rows in each cell are MAE, MSE, Relative MAE, respectively. The notation for the different models is the same as in Table \ref{Tab:ATMHedgingsummary1}.}
\end{table}

\begin{table}[H]
    \centering
    \begin{tabular}{ |p{1.6cm}<{\centering}|p{1.4cm}<{\centering}|p{1.4cm}<{\centering}|p{1.4cm}<{\centering}|p{1.4cm}<{\centering}|p{1.4cm}<{\centering}|p{1.4cm}<{\centering}|p{1.4cm}<{\centering}|p{1.4cm}<{\centering}| }
     \hline
     \multicolumn{9}{|c|}{Next day at-the-money delta hedging for neural network-SDE models} \\
     \hline
     Days & \multicolumn{2}{c|}{NNLV} & \multicolumn{2}{c|}{SDENN} & \multicolumn{2}{c|}{SDENN-Drift} & \multicolumn{2}{c|}{2D-NN} \\
     \hline
      & Call & Put & Call & Put & Call & Put & Call & Put\\
     \hline
     \multirow{3}{*}{2017-09-01}  & 2.892 & 2.784 & 2.922 & 2.821 & 2.797 & 2.761 & 2.482 & 2.458 \\
                                & 9.846 & 9.029 & 10.001 & 9.208 & 9.187 & 8.834 & 7.645 & 7.317  \\
                                & 7.644\% & 12.400\%  & 8.329\% & 12.137\% & 8.460\% & 11.255\% & 5.538\% & 11.443\% \\
     \hline
     \multirow{3}{*}{2017-10-23} & 0.822 & 0.423 & 0.828 & 0.424 & 0.772 & 0.427 & 0.748 & 0.432  \\
                               & 0.916 & 0.282 & 0.924 & 0.285 & 0.808 & 0.275 & 0.757 & 0.269 \\
                               & 4.204\% & 1.738\% & 4.326\% & 1.752\% & 4.282\% & 1.857\% & 3.853\% & 2.003\% \\
     \hline
     \multirow{3}{*}{2017-11-10} & 0.381 & 0.496 & 0.379 & 0.497 & 0.355 & 0.481 & 0.347 & 0.455 \\
                                 & 0.248 & 0.372 & 0.249 & 0.372 & 0.228 & 0.351 & 0.202 & 0.309  \\
                                 & 2.074\% & 3.640\% & 2.070\% & 3.606\% & 2.148\% & 3.413\% & 2.293\% & 3.403\% \\
     \hline
     \multirow{3}{*}{2017-12-07} & 2.424 & 1.284 & 2.432 & 1.229 & 2.234 & 1.142 & 2.260 & 1.071  \\
                                 & 6.715 & 2.187 & 6.601 & 2.035 & 5.696 & 1.728 & 5.584 & 1.713  \\
                                 & 3.090\% & 9.804\% & 3.607\% & 9.355\% & 2.832\% & 8.895\% & 4.344\% & 6.392\% \\
     \hline
     \multirow{3}{*}{2017-01-26} & 7.899 & 2.643  & 7.882 & 2.620 & 7.678 & 2.526 & 7.358 & 2.237  \\
                                 & 63.437 & 8.017 & 63.140 & 7.865 & 59.806 & 7.314 & 54.847 & 5.764  \\
                                 & 3.007\% & 18.608\% & 2.997\% & 18.501\% & 2.926\% & 17.852\% & 2.793\% & 15.803\% \\
     \hline
    \end{tabular}\\
    \caption{Next-day at-the-money delta hedging performance for the neural network models. The three rows in each cell are MAE, MSE, Relative MAE, respectively.  The notation for the different models is the same as in Table \ref{Tab:ATMHedgingsummary2}.}
    % \label{Tab:summary}
\end{table}

We observe that the neural-network-based SDE model underperforms the traditional options pricing models in the case of at-the-money delta hedging.

\subsection{Hedging Out-of-sample Payoff Functions} 
This section describes the hedging performance for the models trained in Section \ref{OutOfSampleC2}. The models are trained only on the call options. Hedging results are provided for both call options (in-sample) and put options (out-of-sample). The following table reports three separate daily results for the next day hedging performance.

\begin{table}[H]
    \centering
    \begin{tabular}{ |p{1.6cm}<{\centering}|p{1.4cm}<{\centering}|p{1.4cm}<{\centering}|p{1.4cm}<{\centering}|p{1.4cm}<{\centering}|p{1.4cm}<{\centering}|p{1.4cm}<{\centering}|p{1.4cm}<{\centering}|p{1.4cm}<{\centering}|}
     \hline
     \multicolumn{9}{|c|}{Hedging performance: train on call options and evaluate on put options.} \\
     \hline
     Days & \multicolumn{2}{c|}{BS-Scalar}  & \multicolumn{2}{c|}{Local Volatility} & \multicolumn{2}{c|}{Heston}  & \multicolumn{2}{c|}{2D-NN} \\
     \hline
      & Call & Put  & Call & Put  & Call & Put  & Call & Put\\
     \hline
     \multirow{3}{*}{2017-09-01} & 0.905 & 1.094 & 5.782 & 2.053 & 1.095 & 1.289 & 0.927 & 1.000\\
                                & 1.584 & 2.810 & 92.507 & 15.588   & 2.201 & 3.504 & 2.108 &2.446 \\
                                & 6.173\% & 17.772\% & 18.058\% & 17.965\%  & 21.047\% & 17.664\% & 3.810\% & 16.696\%\\

    \hline
     \multirow{3}{*}{2017-10-23} & 0.877 & 0.291 & 1.940 & 0.703 & 0.920 & 0.279 & 0.829 & 0.288\\
                                & 1.022 & 0.222 & 9.423 & 2.170   & 1.100 & 0.214 & 0.910 & 0.217\\
                                & 2.761\% & 3.457\% & 5.873\% & 4.005\%  & 5.502\% & 3.500\% & 2.357\% & 3.533\%\\

    \hline
     \multirow{3}{*}{2017-11-10} & 0.350 & 0.279 & 0.996 & 0.458 & 0.366 & 0.304 & 0.386 & 0.269\\
                                & 0.348 & 0.201 & 2.170 & 0.638   & 0.356 & 0.221 & 0.430 & 0.207\\
                                & 2.651\% & 7.782\% & 8.670\% & 7.695\%  & 5.261\% & 7.750\% & 3.250\% & 7.594\%\\

    \hline
    \end{tabular}\\
    \caption{Hedging performance for training on call options and out-of-sample evaluation on put options. There are three rows for each cell. The first row reports MAE, the second row reports MSE, and the third row reports relative MAE. The "BS-Scalar" method uses the Black-Scholes model to estimate the implied volatility (a scalar) for each individual call option and then uses this implied volatility value to calculate the delta for the put option with the same strike price and maturity.}
    % \label{Tab:summary}
\end{table}
%  \vspace{-1.0em}
 We observe that the neural-network-based SDE model outperforms Dupire's local volatility model and the Heston model, but slightly underperforms the Black-Scholes model.

\subsection{Hedging with Recalibration} 
In our final experiment, we explore the hedging performance of the neural network-SDE model in comparison to several benchmark models. We evaluate four models: the Black-Scholes model, the local volatility model, the Heston model, and the two-dimensional neural network-SDE model (2D-NN). The models are calibrated/trained on call options and their hedging performance is calculated for both call and put options. All of the results are for S\&P 500 index options. The models are trained as follows: 

\begin{itemize}
    \item For the Black-Scholes model, we use the implied volatility from the call option for hedging the corresponding put option (at the same strike price $K$ and maturity $T$). Specifically, consider the pair of call and put options on the $i$-th day: $C_{K,T}^i$ and $P_{K,T}^i$. We use the Black-Scholes formula to obtain the implied volatility $\sigma_{K,T}^i$ from the call option price $C_{K,T}^i$. Then, $\sigma_{K,T}^i$ is used to calculate the Black-Scholes delta hedge for the put option $P_{K,T}^i$. $\sigma_{K,T}^i$ is of course also used to calculate the Black-Scholes delta hedge for the call option $C_{K,T}^i$.
    \item The local volatility model is calibrated using Dupire's formula on all of the call options from the entire day.
    \item Both the 2D-NN and Heston model are trained on call options using the stochastic gradient descent method. 
\end{itemize}

Additionally, we consider models with and without recalibration (\ref{Recal}). Specifically, for 2D-NN and Heston models without recalibration, we train the models on the call options for a single day (2017-09-01), and test its hedging performance on call and put options for a subsequent 42 day time period. For models with recalibration, we train the models on the call options on day t and test their hedging performance for both call and put options on day t.

The delta hedge ratio $\Delta_i^t = \frac{\partial P_i^t}{\partial S_t}$ is calculated for each model, where $P_i^t$ is the price of the $i$-th option contract at time $t$ and $S_t$ is the underlying stock price at time $t$. The accuracy of the model-generated delta hedge for a one day time horizon is evaluated in the historical data using the following error metrics: 
\begin{equation}
    \textbf{Mean Absolute Error} = \frac{1}{T} \frac{1}{N} \sum_{t=1}^T \sum_{i=1}^N |\Delta P_i^t - \Delta_i^t \times ( S^{t+1} - S^{t} )|,
\end{equation}
\begin{equation}
    \textbf{Mean Squared Error} = \frac{1}{T} \frac{1}{N} \sum_{t=1}^T \sum_{i=1}^N \bigg{(} \Delta P_i^t - \Delta_i^t \times ( S^{t+1} - S^{t} ) \bigg{)}^2,
\end{equation}
\begin{equation}
    \textbf{Relative Mean Absolute Error} = \frac{1}{T} \frac{1}{N} \sum_{t=1}^T \sum_{i=1}^N \frac{ |\Delta P_i^t - \Delta_i^t \times ( S^{t+1} - S^{t} )| }{ P_i^t} \times 100\%,
\end{equation}
where $\Delta P_i^t = P_i^{t+1} - P_i^t$ is the actual price change of the contract $i$ and $S^t$ is the underlying stock price at time $t$. A finite-difference formula is applied to two Monte Carlo simulations to calculate the delta hedge $\Delta_i^t$ for the Heston, local volatility, and neural network-SDE models.

The following two tables present the hedging performance with and without recalibration for a two month time period. In both tables, the 2D neural network models achieve lower error for both call and put options according to all evaluation metrics. The Heston model and the local volatility model do not improve with recalibration; however, recalibration improves the hedging performance of the 2D neural network model. 

\begin{table}[H]
    \centering
    \begin{tabular}{ |p{2.0cm}<{\centering}|p{1.3cm}<{\centering}|p{1.3cm}<{\centering}|p{1.3cm}<{\centering}|p{1.3cm}<{\centering}|p{1.3cm}<{\centering}|p{1.3cm}<{\centering}|p{1.3cm}<{\centering}|p{1.3cm}<{\centering}| }
     \hline
     \multicolumn{9}{|c|}{Hedging for two months period (without recalibration).} \\
     \hline
     Days & \multicolumn{2}{c|}{BS}  & \multicolumn{2}{c|}{Local Volatility} & \multicolumn{2}{c|}{Heston}  & \multicolumn{2}{c|}{2D-NN} \\
     \hline
      & Call & Put  & Call & Put  & Call & Put  & Call & Put\\
     \hline
     \multirow{3}{*}{\makecell{2017/09 $\sim$\\ 2017/10}} & NA & NA & 2.466 & 0.870 & 1.044 & 0.486 & 0.963 & 0.449\\
                                & NA & NA & 19.332 &  4.171  & 2.138 & 0.674 & 1.875 & 0.615\\
                                & NA & NA & 7.698\% & 7.344 \%  & 7.224\% & 6.982\% & 2.348\% & 6.650\%\\

    \hline
    \end{tabular}\\
    \caption{Out-of-sample hedging performance over a two month time period period without recalibration. There are three rows for each cell. The first row reports MAE, the second row reports MSE, and the third row reports relative MAE.}
    \label{Tab:HedgeOutofSample1}
\end{table}

\begin{table}[H]
    \centering
    \begin{tabular}{ |p{2.0cm}<{\centering}|p{1.3cm}<{\centering}|p{1.3cm}<{\centering}|p{1.3cm}<{\centering}|p{1.3cm}<{\centering}|p{1.3cm}<{\centering}|p{1.3cm}<{\centering}|p{1.3cm}<{\centering}|p{1.3cm}<{\centering}| }
     \hline
     \multicolumn{9}{|c|}{Hedging for two months period (with recalibration).} \\
     \hline
     Days & \multicolumn{2}{c|}{BS}  & \multicolumn{2}{c|}{Local Volatility} & \multicolumn{2}{c|}{Heston}  & \multicolumn{2}{c|}{2D-NN} \\
     \hline
      & Call & Put  & Call & Put  & Call & Put  & Call & Put\\
     \hline
     \multirow{3}{*}{\makecell{2017/09 $\sim$\\ 2017/10}} & 0.986  & 0.473 & 2.813 & 0.950 & 1.043 & 0.487 & 0.947 & 0.428\\
                                & 1.999 & 0.635 & 31.715 & 5.811 & 2.139 & 0.675 & 1.781 & 0.557\\
                                & 2.887\% & 6.967\% & 7.696\% &  7.540\%  & 6.823\% & 6.995\% & 2.876\% & 6.747\%\\

    \hline
    \end{tabular}\\
    \caption{Out-of-sample hedging performance over a two month time period with recalibration. The evaluation metrics are the same as in the previous table.}
    \label{Tab:HedgeOutofSample2}
\end{table}

%Bibliography
\bibliographystyle{unsrt}  
\bibliography{references}

\end{document}